\documentclass[twocolumn]{aastex631}

\received{November 29, 2025}
\revised{February 26, 2026}
\accepted{March 5, 2026}

\submitjournal{ApJ}

\begin{document}

\title{The Cluster Evolutionary Reference Ensemble at Low-$z$ (CEREAL) Sample of Galaxy Clusters I: X-ray Morphological Properties and Demographics}

\author[0000-0001-6699-1300]{Laurel White}
\affiliation{Massachusetts Institute of Technology \\
77 Massachusetts Avenue \\
Cambridge, MA 02139, USA}

\author[0000-0001-5226-8349]{Michael McDonald}
\affiliation{Massachusetts Institute of Technology \\
77 Massachusetts Avenue \\
Cambridge, MA 02139, USA}

\author[0000-0003-0667-5941]{Steven W. Allen}
\affiliation{Kavli Institute for Particle Astrophysics and Cosmology, Stanford University \\ 452 Lomita Mall \\ Stanford, CA 94305, USA}
\affiliation{Department of Physics, Stanford University \\ 382 Via Pueblo Mall \\ Stanford, CA 94305, USA}
\affiliation{SLAC National Accelerator Laboratory \\ 2575 Sand Hill Road \\ Menlo Park, CA 94025, USA}

\author[0000-0002-1379-4482]{Marshall W. Bautz}
\affiliation{MIT Kavli Institute for Astrophysics and Space Research \\ Massachusetts Institute of Technology \\ Cambridge, MA 02139, USA}

\author[0000-0002-2238-2105]{Michael Calzadilla}
\affiliation{Harvard-Smithsonian Center for Astrophysics \\ 60 Garden Street \\ Cambridge, MA 02138, USA}

\author[0000-0002-7371-5416]{Gordon P. Garmire}
\affiliation{Huntingdon Institute for X-ray Astronomy, LLC \\ 10677 Franks Rd \\ Huntingdon, PA 16652, USA}

\author[0000-0001-7271-7340]{Julie Hlavacek-Larrondo}
\affiliation{Département de Physique, Université de Montréal \\ Succ. Centre-Ville \\ Montréal, Québec, H3C 3J7, Canada}

\author[0000-0002-0765-0511]{Ralph Kraft}
\affiliation{Harvard-Smithsonian Center for Astrophysics \\ 60 Garden Street \\ Cambridge, MA 02138, USA}

\author[0000-0002-8031-1217]{Adam B. Mantz}
\affiliation{Kavli Institute for Particle Astrophysics and Cosmology, Stanford University \\ 452 Lomita Mall \\ Stanford, CA 94305, USA}

\author[0000-0003-3521-3631]{Taweewat Somboonpanyakul}
\affiliation{Department of Physics, Faculty of Science, Chulalongkorn University \\ 254 Phyathai Road \\ Patumwan, Bangkok 10330, Thailand}

\author[0000-0001-8121-0234]{Alexey Vikhlinin}
\affiliation{Harvard-Smithsonian Center for Astrophysics \\ 60 Garden Street \\ Cambridge, MA 02138, USA}



\begin{abstract}

With rapid improvements in the assembly of large samples of galaxy clusters, we are approaching the ability to study clusters at $z\gtrsim2$. Evolutionary studies comparing these distant clusters to the clusters in our local universe depend heavily on the reliability of low-redshift cluster samples, most of which are subject to X-ray selection effects, biasing them to relaxed, cool core clusters. Here, we introduce the Cluster Evolutionary Reference Ensemble At Low-$z$ (CEREAL) sample, composed of \emph{Chandra} X-ray observations of 169 galaxy clusters that have been selected from the \emph{Planck} Sunyaev-Zel'dovich catalog. CEREAL has a simple and well-understood selection function, spans an order of magnitude in mass at $z\sim0.15$, and has uniform, high-resolution X-ray follow-up. We present the full sample and provide results based on X-ray surface brightness properties, finding significantly more non-cool core systems than in X-ray-selected samples. We use surface brightness concentration (c$_\mathrm{SB}$) as a proxy for cool core strength and centroid shift ($w$) to measure dynamical state. Over the full sample, we find a cool core (c$_\mathrm{SB} > 0.075$) fraction of $0.39_{-0.04}^{+0.04}$, a strong cool core (c$_\mathrm{SB} > 0.155$) fraction of $0.13_{-0.03}^{+0.03}$, and a dynamically relaxed ($w<0.01$) fraction of $0.42_{-0.04}^{+0.04}$. We find no mass dependence in the fraction of clusters that appear relaxed or have cool cores. We quantify the rarity of X-ray-bright central point sources (L$_\mathrm{nuc,~2-10~keV} > 10^{43}$ erg s$^{-1}$), finding them to be intrinsically rare ($0.7_{-0.5}^{+1.2}$\% of massive, low-z clusters) with a notable increase in occurrence rate at the centers of cool cores.

\end{abstract}



\section{Introduction} \label{sec:intro}

Galaxy clusters are dense environments that provide unique insights into the evolution of galaxies and the cosmic web. As the most massive collapsed structures, galaxy clusters provide maximal leverage for cosmological studies targeting the growth of structure \citep[e.g.,][]{2002ApJ...573....7E}, while also  serving as ``closed box" models of the universe in which we can probe galaxy formation, baryon cycling, and chemical enrichment \citep[e.g.,][]{2012ARA&A..50..353K}. In their centers, galaxy clusters host the most massive black holes in the universe, facilitating some of the most detailed studies of the effects of black hole feedback on the surrounding matter \citep[e.g.,][]{2007ARA&A..45..117M}.

To date, X-ray observations of galaxy clusters extending out to $z\sim1.5$ reveal relatively little evolution in the properties of the intracluster medium \citep[hereafter ICM;][]{2010A&A...521A..64S, 2012A&A...539A.105S, 2013ApJ...774...23M, 2015MNRAS.447.3723P, 2016MNRAS.456.4020M, 2017ApJ...843...28M, 2018MNRAS.474.1065S, 
2021ApJ...910...14G, 2023ApJ...948...49R}, indicating that the properties of the ICM were established during or shortly after cluster formation and virialization. These studies are generally limited by the combination of the small number of known clusters at high redshift and the faintness of these distant systems, which require expensive observations. One promising and largely unexplored path forward in understanding cluster evolution is to improve our understanding of clusters in the nearby universe, reducing the uncertainty in the evolution by precisely anchoring the low-redshift state. While there have been extensive studies of many nearby clusters \citep[e.g.,][]{2004A&A...425..367B, 2007A&A...469..363B, 2007ApJS..172..561B}, these have primarily used cuts in X-ray flux to select clusters, which can be inherently biased towards relaxed clusters \citep{2017MNRAS.468.1917R}. This is because relaxed systems tend to have strongly cooling, dense cores, and the X-ray flux scales with the square of the gas density.

\begin{figure*}[ht!]
\centering
\includegraphics[width=0.99\textwidth]{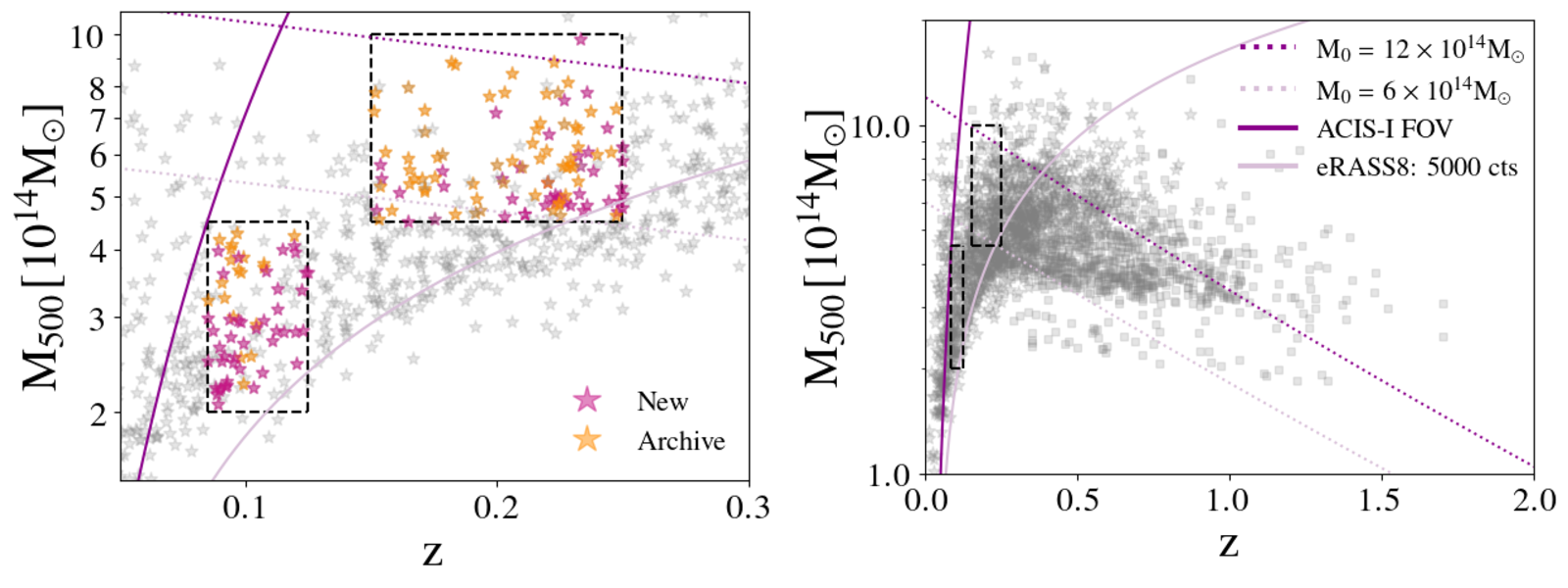}
\caption{The mass-redshift distribution of clusters in the CEREAL sample. Left panel: The grey stars are clusters selected via the \textit{Planck} SZ survey \citep{2016A&A...594A..27P}. Colored points are clusters that are included in the CEREAL sample, where the point color refers to whether or not new observations were obtained as part of this program. Right panel: Grey stars are \textit{Planck} clusters and grey squares are SPT clusters \citep{2015ApJS..216...27B}, showing a representative distribution of SZ-selected clusters. In both panels, the dotted purple line represents the evolution of a cluster with a mass like that of Coma, while the dotted pink line corresponds to a cluster like Perseus. The solid pink line represents the cutoff above which clusters will be observed with at least 5000 counts in the eRosita All-Sky Survey (eRASS8), and the solid purple line represents the field of view of ACIS-I on \textit{Chandra}. These lines motivate the mass and redshift range of the CEREAL sample.}
\label{fig:evol}
\end{figure*}

One method of selecting galaxy clusters that may be less biased to dynamical state \citep{2015ApJ...802...34L,2024MNRAS.534.2378K} relies upon the integrated Sunyaev-Zel'dovich (SZ) effect \citep{1972CoASP...4..173S}. The hot plasma in galaxy clusters causes inverse Compton scattering of photons from the Cosmic Microwave Background (CMB), leaving ``shadows" imprinted on CMB maps that appear as deficits in photon intensity at frequencies less than 218\,GHz. The strength of the SZ signal depends on the integrated gas pressure through the cluster, which is proportional to the total mass of the cluster. Therefore, implementing a cut based on the strength of the SZ signal is similar to implementing a mass cut. Further, since the integrated gas pressure scales with density (rather than the square of density like X-ray flux), this selection method is less biased towards relaxed, cool core systems. We take advantage of this fact in compiling a well-defined sample of low-redshift galaxy clusters: the Cluster Evolutionary Reference Ensemble at Low-z (CEREAL).

The value of such a sample can be highlighted by considering previous claims of a bimodality in the core properties of galaxy clusters, between cool cores and non-cool cores \citep[e.g.,][]{2009ApJS..182...12C,2010A&A...513A..37H}. Subsequent works, based on less biased and better-understood samples, have found this bimodality to be a selection effect, and that the distribution of core properties is more likely a continuum, with many more non-cool cores and a significant population of ``moderate'' cool core systems \citep[e.g.,][]{2017MNRAS.468.1917R, 2021ApJ...914...58A, 2024A&A...687A.129L}. Through a systematic imaging analysis using observations with a uniform minimum depth, we aim to use the CEREAL sample to cleanly constrain low-redshift cluster demographics such as the cool core fraction, the dynamical state, and the frequency of X-ray-bright central point sources. In future work, we will examine thermodynamic profiles.

This paper is laid out as follows. In \S\ref{sec:selection}, we describe the selection of the CEREAL sample. In \S\ref{sec:methods}, we apply various imaging analyses to our complete set of \textit{Chandra} observations, and we show the population statistics (cool core fraction, relaxed fraction, and central point source fraction) in \S\ref{sec:results}. In \S\ref{sec:discussion}, we discuss the implications of our findings, including the completeness of the CEREAL sample compared to other samples. We conclude in \S\ref{sec:conclusion}. Throughout, we use a standard $\Lambda$CDM cosmology with H$_0$ = 70 km s$^{-1}$ Mpc$^{-1}$, $\Omega_{\rm{M}}$ = 0.3, and $\Omega_{\Lambda}$ = 0.7. This work employs a list of \textit{Chandra} datasets, obtained by the \textit{Chandra X-ray Observatory}, contained in the Chandra Data Collection (CDC) 482~\dataset[DOI: 10.25574/cdc.482]{https://doi.org/10.25574/cdc.482}.

\begin{figure*}[ht!]
\centering
\includegraphics[width=0.9\textwidth]{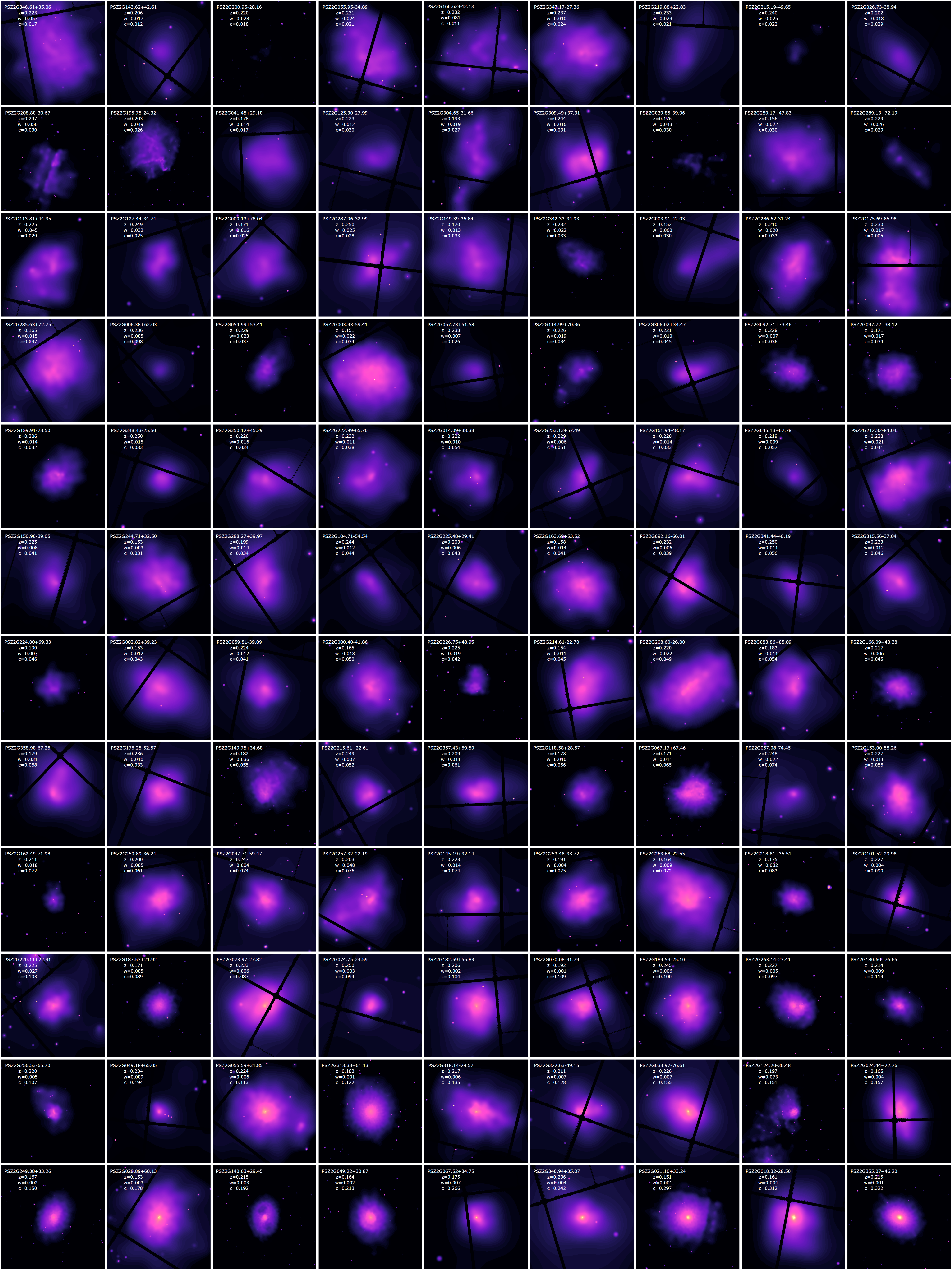}
\caption{\textit{Chandra} images of all high-mass CEREAL clusters. Clusters are shown on 550$\times$550 pixel grids and with the same rest-frame intensity coloring, with $(1+z)^{-4}$ surface brightness dimming divided out, meaning that clusters that appear fainter in this figure \emph{are} intrinsically less luminous or lower surface brightness. Images are adaptively smoothed to improve visual clarity for a sample spanning a wide range in signal-to-noise. For clusters with a single observation, ACIS-I chip gaps and regions with low exposure ($<$50\%) are shown in black. This is done for illustrative purposes only, and a more careful exposure correction is done prior to analysis. Clusters are labeled with their \textit{Planck} names and redshifts as well as their centroid shift ($w$) and surface brightness concentration values ($c$; calculated using 40 kpc and 400 kpc apertures) as described in \S\ref{sec:methods}.}
\label{fig:highm_img}
\end{figure*}

\begin{figure*}[ht!]
\centering
\includegraphics[width=0.9\textwidth]{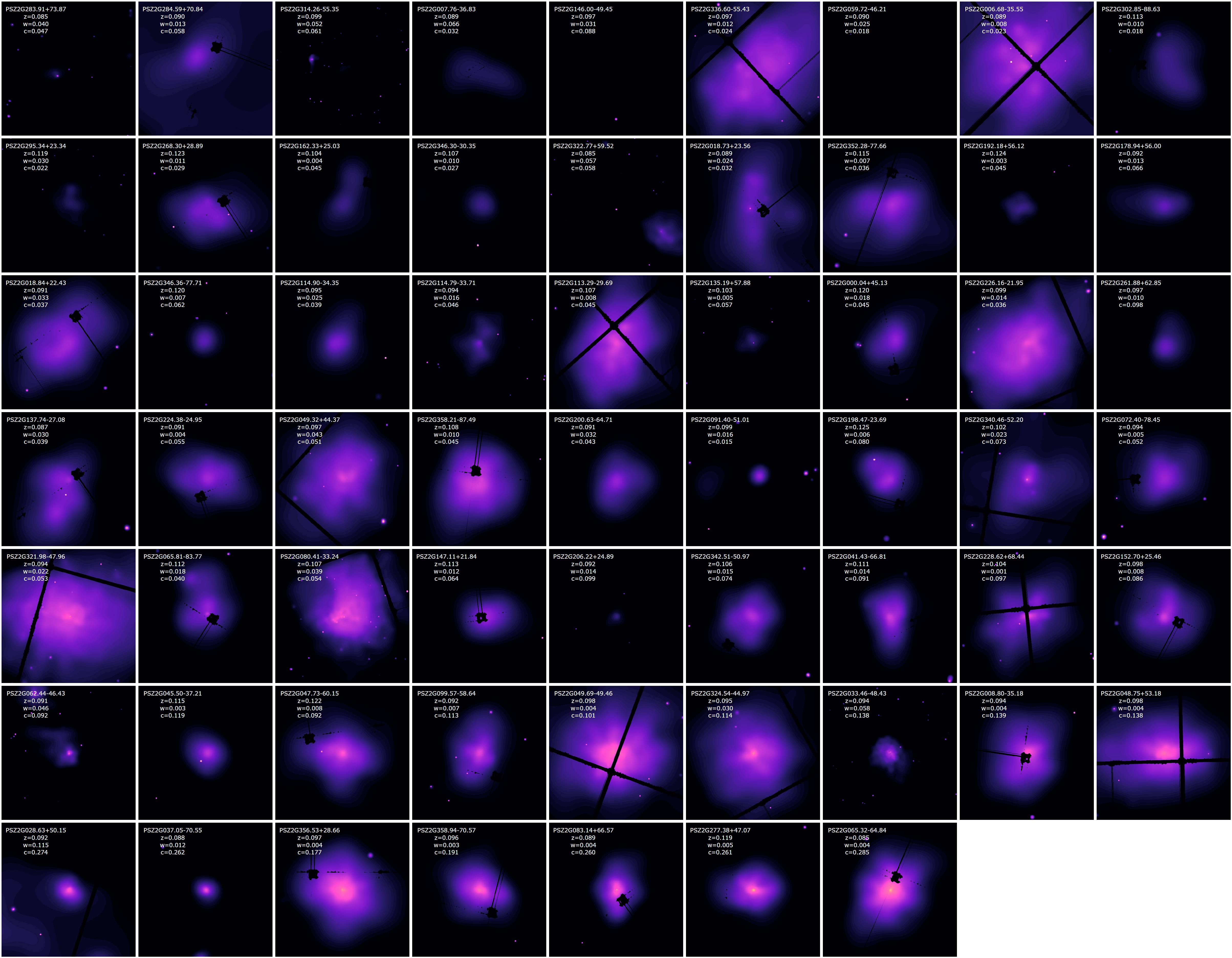}
\caption{Similar to Figure \ref{fig:highm_img}, but for the low-mass CEREAL clusters. Both sets of images use the same scaling for comparative purposes.}
\label{fig:lowm_img}
\end{figure*}

\section{The CEREAL Sample} \label{sec:selection}

The current frontier of galaxy cluster science is beyond $z\sim2$, closing in on the era of formation of the first massive clusters. In particular, modern SZ surveys like the one being carried out by the third-generation detector on the South Pole Telescope \citep[SPT-3G;][]{2014SPIE.9153E..1PB} have dramatically increased sensitivity compared to previous generations. They are capable of probing the progenitor population of the most massive nearby clusters as they would have likely appeared shortly after their collapse. To fully exploit these efforts, we need large, well-understood samples of nearby clusters that are the descendants of these massive clusters being discovered at early times.

With this in mind, we have compiled the CEREAL sample. The clusters in this sample are selected from the \textit{Planck} 2nd SZ survey \citep{2016A&A...594A..27P} in two regimes, as shown in Figure \ref{fig:evol}: a high-mass regime where the M$_{500}$ values fall between the evolutionary tracks of the Perseus-like cluster and the Coma-like cluster, and a low-mass regime consisting of clusters at similar redshift values but with masses a factor of two lower. We define the evolutionary tracks of massive galaxy clusters following \cite{2010MNRAS.406.2267F}. The high-mass clusters will be, on average, the descendants of the massive clusters that will be discovered by SPT-3G at $z>2$, while the other sample will more closely match the high-z clusters in actual mass, making the CEREAL sample a good anchor for comparative evolutionary studies and allowing us to determine how various properties trend with both mass and redshift. While these two samples are slightly offset in redshift (due to limitations in \emph{Planck} sensitivity), we emphasize that the difference in cosmic time between the two subsamples is only $\sim$1\,Gyr. In contrast, an average halo in the low-mass sample would require $\sim$4\,Gyr to evolve into the average cluster in the high-mass sample. So, we take this difference in observed redshift to be subdominant in the evolutionary state of these systems.

We note that low-redshift clusters are well-studied and that the CEREAL sample enters a landscape containing many other samples. The existing samples can be broadly categorized into X-ray selected and SZ-selected samples. The X-ray selected samples include, for example, the MAssive Cluster Survey \citep[MACS;][]{2001ApJ...553..668E}, the HIghest X-ray FLUx Galaxy Cluster Sample \citep[\textit{HIFLUGCS;}][]{2002ApJ...567..716R}, the 400 Square Degree sample \citep[400 SD;][]{2007ApJS..172..561B}, the Representative \textit{XMM-Newton} Cluster Structure Survey \citep[REXCESS;][]{2007A&A...469..363B}, work by \cite{2008A&A...483...35S}, the \textit{ACCEPT} sample \citep{2009ApJS..182...12C}, work by \cite{2015MNRAS.449..199M}, work by \cite{2016ApJ...819...36D}, work by \cite{2017ApJ...841....5N}, work by \cite{2020MNRAS.497.5485Y}, and the eROSITA Final Equatorial Depthy Survey \citep[eFEDS;][]{2022A&A...661A..12G}. They are primarily selected using minimum X-ray flux cuts, with some samples being limited to only the most massive---and therefore brightest---clusters and some covering a wider, more representative mass range. The SZ-selected samples include, for example, \textit{Chandra} follow-up of SPT-selected clusters by \cite{2012ApJ...761..183S}, \cite{2013ApJ...774...23M}, and \cite{2017ApJ...841....5N}, \textit{Chandra} follow-up of \textit{Planck}-selected clusters by \cite{2017ApJ...843...76A,2021ApJ...914...58A}, and the Cluster HEritage project with \textit{XMM-Newton} - Mass Assembly and Thermodynamics at the Endpoint of structure formation \citep[CHEX-MATE;][]{2021A&A...650A.104C}.

The advantages of the CEREAL sample are that (i) it is selected based on the SZ signal, which is less biased to the cluster dynamical state than X-ray selection and provides a more direct comparison to high-$z$ cluster samples which are predominantly selected in the SZ, and (ii) it has a well-understood selection function, being cleanly defined within a narrow range in mass and redshift. In particular, the CEREAL sample is an excellent complement to the CHEX-MATE sample and the work done by \cite{2017ApJ...843...76A,2021ApJ...914...58A}, both of which are also SZ-selected and well-defined. The latter focuses on the most massive clusters out to $z\sim 0.35$, whereas the former is divided into low-mass (Tier 1) and high-mass (Tier 2) subsamples spanning different redshift ranges. Whereas these previous samples cover a wider redshift range, the CEREAL sample comprises a large ($>100$) number of clusters in a narrow redshift band but spanning a wide mass range.

The CEREAL sample is defined based on a series of four cuts, making a rectangle in mass-redshift space. The lower redshift cut of $z>0.15$ ($z>0.085$ for the low-mass sample) is chosen for several reasons. First, we must look far enough in distance such that the volume of the universe is sufficiently large that it contains a significant number of massive of galaxy clusters. Further, clusters must be far enough away that their apparent size is smaller than the \textit{Chandra} ACIS-I field of view. The apparent size of a cluster is assumed to be R$_{500}$, which can be calculated directly from the SZ-derived value of M$_{500}$. Lastly, clusters at $z > 0.15$ are unresolved to \textit{Planck}, so the SZ signal will not be spread across multiple beams but rather matched in apparent size to a single beam. This ensures a clean selection and roughly matches the apparent beam-to-cluster size ratio for a massive cluster at $z>1$ as observed by SPT.
The upper redshift cut at $z < 0.25$ ($z < 0.125$ for the low-mass sample) is chosen such that the full-depth eROSITA All-Sky Survey eRASS8 will obtain, on average, 5000 counts per cluster \citep{2012MNRAS.422...44P}. This will facilitate a future extension to this work that will combine Chandra data on small scales and eROSITA data on large scales. We note that the eROSITA mission has thus far completed four complete scans of the sky, yielding approximately half as many counts as projected for the full-depth survey. We make only one additional cut, which requires that the clusters lie more than 20$^{\circ}$ from the Galactic plane. 

We obtain \emph{Chandra} observations for every cluster within these mass and redshift cuts. For clusters that satisfy these criteria and do not have archival \emph{Chandra} ACIS-I data, we obtain short new exposures (5--12\,ks). At higher masses (M$_{500} = 4.5-10.0\times10^{14}$\,M$_{\odot}$), this selection yields 61 clusters with archival \textit{Chandra} observations and 47 for which we have obtained new \textit{Chandra} observations (Chandra program 23800272), giving an overall high-mass sample of 108 clusters (see Figure \ref{fig:highm_img}). At lower masses (M$_{500} = 2.0-4.5\times10^{14}$\,M$_{\odot}$), we have 19 clusters with archival observations and 44 which we have newly observed (Chandra program 25800481) for a total of 63 low-mass clusters. We find that 2 of the low-mass clusters---PSZ2G127.71-69.55 and PSZ2G255.07+54.84---are not detected by \textit{Chandra} despite high-significance detections by \textit{Planck}. These may be mis-classified in redshift (leading to X-ray luminosity being over-predicted), false detections, or systems for which the X-ray surface brightness is significantly lower than predicted for other reasons. Referring to the Planck catalog, PSZ2G127.71-6955 was flagged as likely being a false detection with $>$99\% confidence by their neural network quality assessment algorithm. On the other hand, PSZ2G255.07+54.84 has an optical counterpart with a secure spectroscopic redshift and a high signal-to-noise ($>$5) in the Planck SZ catalog. This system, given its low SZ-derived mass, may be the line-of-sight projectection of smaller structures, or a very low X-ray surface brightness system.
These two systems are removed from the sample, leaving us with 61 low-mass clusters (see Figure \ref{fig:lowm_img}) and 169 total clusters in the sample.

The final sample of 169 clusters has a well-understood selection function and avoids the biases of X-ray- and optically-selected samples. Each cluster has been observed for sufficient time to obtain at least 2000 total counts in the full energy band, allowing for a uniform analysis to be applied across the full sample. Count rates were estimated using the L$_X$-M$_\mathrm{tot}$ relation from \cite{2020ApJ...892..102L}. Note that these clusters are nearby and massive enough that background count rates are minimal. Finally, while any biases that are inherent to SZ selection will apply to this sample, they will also apply to the samples that CEREAL is designed for comparison to -- namely high-redshift, SZ-selected cluster samples.

\section{Methods} \label{sec:methods}
The goal of this paper is to understand the demographics of low-$z$ clusters, focusing on quantities that can be derived from X-ray surface brightness. In subsequent papers we will present the thermodynamic properties (which require spectroscopic analysis) and multi-wavelength properties.
We make use of the same analysis pipeline described in \cite{2025ApJ...988...24W}. The data processing relies upon the standard tools provided in version 4.17 of the Chandra Interactive Analysis of Observations (CIAO) software package as well as the CALDB 4.12 calibration files \citep{2006SPIE.6270E..1VF}.

\begin{figure}[t!]
\includegraphics[width=0.49\textwidth]{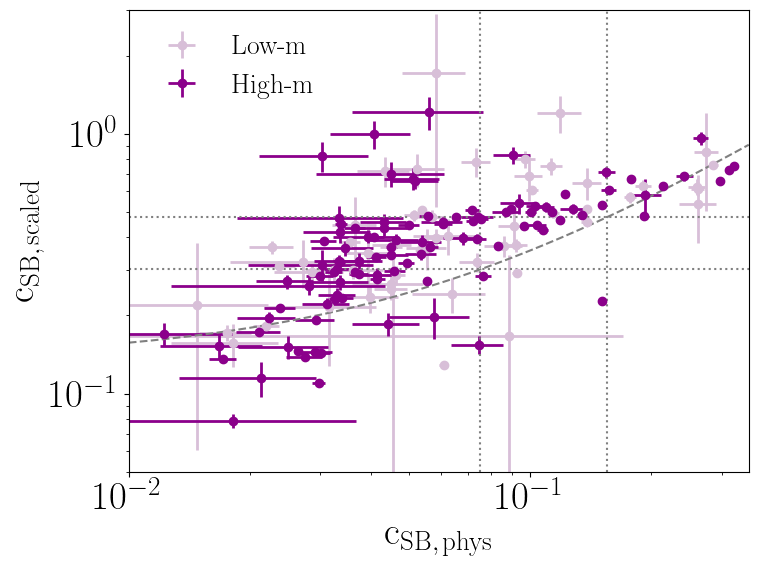}
\caption{A comparison of the concentration values measured using physical apertures (40 kpc and 400 kpc) versus using fractions of the radius ($0.15R_{500}$ and $R_{500}$). Significant outliers have been cropped for visualization purposes. The dashed line represents the line of best fit. The dotted lines represent the cutoffs for moderate and strong cool cores.
\label{fig:csb}}
\end{figure}

\begin{figure*}[t!]
\plotone{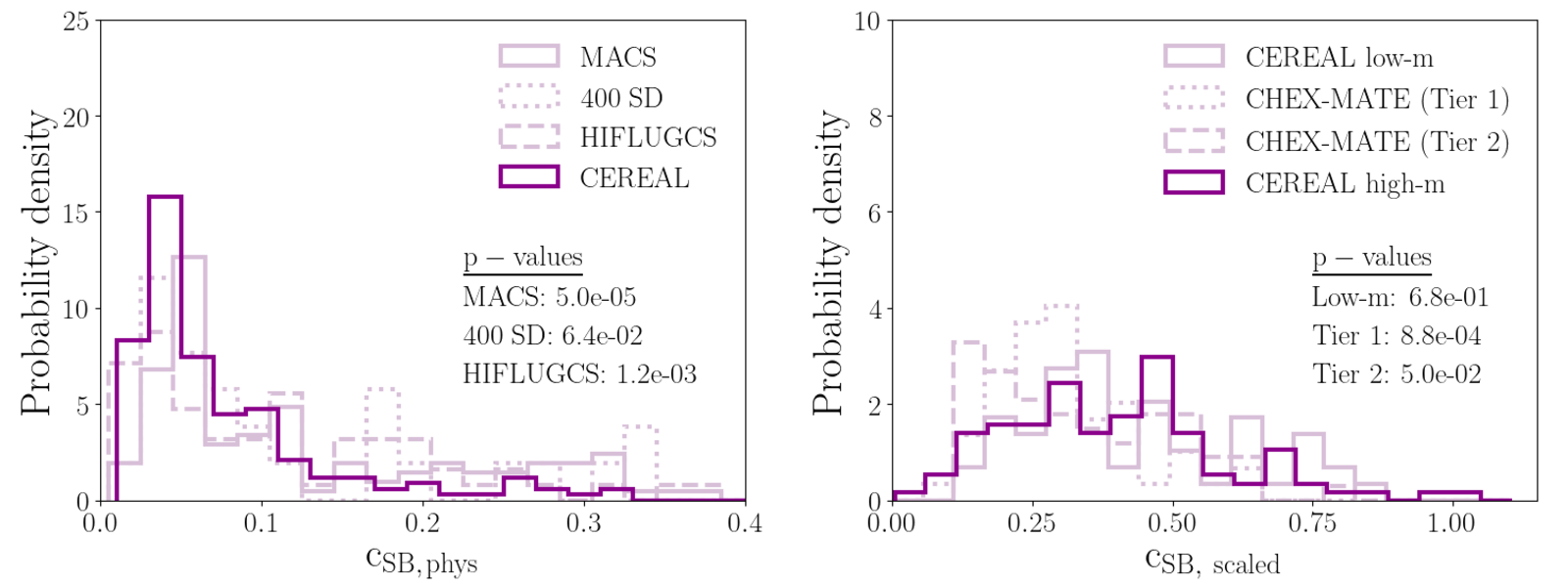}
\caption{Left: The distributions of concentration values for the CEREAL sample, the MACS sample \citep{2017MNRAS.468.1917R}, the 400 SD Sample \citep{2010A&A...521A..64S}, and the \textit{HIFLUGCS} sample \citep{2010A&A...513A..37H}, calculated using physical apertures of 40 kpc and 400 kpc. Right: The distributions of concentration values for the CEREAL mass subsamples and the CHEX-MATE subsamples, calculated using apertures that scale with radius. Each plot shows the p-value for the null hypothesis that the measured distributions are drawn from the same underlying distribution (the left plot compares all distributions to CEREAL, and the right plot compares all distributions to CEREAL high-m), calculated using a K-S test. There is a statistically significant difference between the CEREAL sample and the MACS,  and \textit{HIFLUGCS} samples, while no significant difference is found between the CEREAL and 400 SD samples. Similarly, the low-m and high-m CEREAL subsamples are consistent with each other, and the CEREAL high-m sample is consistent with the CHEX-MATE Tier 2 subsample but not Tier 1.
\label{fig:conc}}
\end{figure*}

\subsection{Data reduction}
We first reprocess the data using \texttt{chandra\_repro}\footnote{\url{https://cxc.cfa.harvard.edu/ciao/ahelp/chandra_repro.html}} according to the most recent calibration files available at the time of the analysis (provided in CALDB version 4.12). This applies the latest charge transfer ineffiency (CTI) correction, time-dependent gain adjustment, and gain map, as well as pulse height amplitude randomization, a sub-pixel adjusment, and cleaning of the VFAINT background.

We then use the \texttt{wvdecomp} routine from the ZHTOOLS package (based on the methodology described in \citealt{1998ApJ...502..558V}) to identify point sources and \texttt{deflare}\footnote{\url{https://cxc.cfa.harvard.edu/ciao/ahelp/deflare.html}} to remove any time intervals containing $\geq2\sigma$ flares. Finally, we use \texttt{merge\_obs}\footnote{\url{https://cxc.cfa.harvard.edu/ciao/ahelp/merge_obs.html}} to combine all observations of a cluster and produce a flux image in the 0.5-7.0 keV band.

\subsection{Surface brightness concentration}
Some of the different methods of determining whether or not a cluster is a cool core are discussed in \cite{2010A&A...513A..37H}. One quantity of interest is the surface brightness concentration\footnote{We will refer to this quantity as ``concentration" throughout this work. It is distinct from the concentration parameter related to cluster density profiles.}, \textit{c$_{\mathrm{SB}}$}, which \citet{2008A&A...483...35S} defined as
\begin{equation}
     c_{\mathrm{\tiny SB}} = \frac{\rm{Flux (r<40 kpc)}}{\rm{Flux (r<400 kpc)}}.
 \end{equation}
While this quantity does not include a measurement of temperature and therefore cannot technically identify cool cores, it has been shown to correlate strongly with central X-ray cooling time and temperature \citep{2010A&A...513A..37H}. In particular, the presence of a cool core in a galaxy cluster corresponds to a high gas density at the center, and X-ray flux is proportional to the square of gas density, so cool cores tend to have centrally peaked fluxes. Therefore, we will use the term ``cool core” throughout when referring to highly-concentrated cores.

Generally, higher values of $c_{\mathrm{\tiny SB}}$ are associated with cool cores and lower values are associated with non-cool cores. \citet{2008A&A...483...35S} used their distribution of $c_{\mathrm{\tiny SB}}$ values for both low-z and high-z samples to create three categories of clusters: strong cool cores with $c_{\mathrm{\tiny SB}} > 0.155$, moderate cool cores with $0.075 < c_{\mathrm{\tiny SB}} < 0.155$, and non-cool cores with $c_{\mathrm{\tiny SB}} < 0.075.$ In \cite{2010A&A...513A..37H}, it was demonstrated that each of their systems with $c_{\rm{SB}} > 0.155$ had a central cooling time below 1 Gyr, validating this statistic as an excellent proxy for cooling state in the absence of spectroscopic information. We calculate the concentration on the 0.5-7.0 keV exposure-corrected flux image of each cluster in units of photons cm$^{-2}$ s$^{-1}$, using the X-ray peak as the cluster center, and using the \texttt{dmstat}\footnote{\url{https://cxc.cfa.harvard.edu/ciao/ahelp/dmstat.html}} routine to sum the fluxes in the 40 kpc and 400 kpc aperture regions. The X-ray peak is determined using \texttt{dmstat} to identify the brightest pixel in a heavily smoothed (15$\sigma$) version of the flux image with point sources removed. We implement a background subtraction that estimates and subtracts off the expected background flux based on local (on-chip) background regions defined beyond $R_{500}$. In addition, we remove point sources, filling these regions with the local average value using \texttt{dmfilth}. \footnote{\url{https://cxc.cfa.harvard.edu/ciao/ahelp/dmfilth.html}}

It is also common to define a concentration value using apertures which scale with the size of the cluster, as in \cite{2022A&A...665A.117C}. From here on out, we will refer to the original value, using the physically-sized 40 kpc and 400 kpc apertures, as $c_{\rm{SB,~phys}}$, and we calculate a new value,
\begin{equation}
    c_{\mathrm{\tiny SB,~scaled}} = \frac{\rm{Flux (r<0.15 R_{500})}}{\rm{Flux (r< R_{500})}}.
\end{equation}
We compare the two methods in \S\ref{sec:results}.

\subsection{Centroid shift}

As a measure of morphological asymmetry, we employ the centroid shift, which is a measure of how stable the X-ray centroid is for a series of annuli spanning the core to the outskirts. The centroid shift, \textit{w}, is defined by \cite{2010A&A...514A..32B} as

\begin{equation}
    w = \left[ \frac{1}{N-1} \sum (\Delta_i-\langle \Delta\rangle)^2 \right]^{1/2} \times \frac{1}{R_{500}},
\end{equation}

\noindent where $N$ is the number of apertures used, $\Delta_i$ is the distance between the peak of the X-ray emission and the centroid of the emission calculated using the \textit{i}th aperture, and $\langle \Delta \rangle$ is the average of these distances. We calculate the centroid shift on the exposure-corrected flux image of each cluster with point sources removed as above. We use 10 circular apertures, centered on the X-ray peak, with radii evenly spaced from 0.1R$_{500}$ out to R$_{500}$. We use the \texttt{dmstat} routine to locate the X-ray peak as well as the centroid of each of the 10 regions for the exposure-corrected flux images. In \cite{2010A&A...514A..32B}, they note a natural gap in the centroid shift distribution around $w\sim 0.01$, so we use this as the threshold between relaxed and disturbed clusters, with symmetric (relaxed) clusters tending to have lower values of $w$ and asymmetric (disturbed) clusters having higher values of $w$. We emphasize that $w\sim 0.01$ is not a physically significant value and cannot be used to definitively delineate between relaxed and disturbed clusters, but it is useful for comparative purposes.

\subsection{Nuclear luminosity}

Following the discovery of the central X-ray-bright point source in a CEREAL cluster (Abell 1885; \citealt{2025ApJ...988...24W}), we are motivated to quantify the occurrence rate of these point sources, which can provide insight into the duty cycle of black hole accretion in the centers of galaxy clusters. Similar to the methodology put forth in \cite{2013MNRAS.431.1638H}, we estimate the degree to which the emission is centrally peaked by measuring the nuclear luminosity at the peak of the cluster emission. We do this by measuring the excess count rate in the 0.5-7.0 keV band in a 1 arcsecond region centered on the X-ray peak and then the background count rate in a 1 to 5 arcsecond annulus also centered on the peak. The background count rate, corrected for the ratio of the areas, is subtracted from the central count rate to obtain an excess count rate. This excess count rate is then converted to a 2-10 keV flux using the web-based PIMMS tool,\footnote{\url{https://cxc.harvard.edu/toolkit/pimms.jsp}} which provides a cycle-dependent conversion factor. This is necessary because the sensitivity of \textit{Chandra} in the soft X-ray band degrades year over year \footnote{\url{https://cxc.cfa.harvard.edu/ciao/why/acisqecontamN0015.html}}, so a different exposure time is needed for each observing cycle in order to obtain the same total flux. We assume a power law model with an index of 1.8 (the average for AGN; \citealt{2006A&A...451..457T}) and an intrinsic column density of 0.01 $\times 10^{22}\mathrm{~cm}^{-2}$. Clusters with multiple observations use a weighted average of the conversion factors. From this flux, and the measured redshift, we obtain the 2-10 keV nuclear luminosity.

\begin{figure}[b]
\plotone{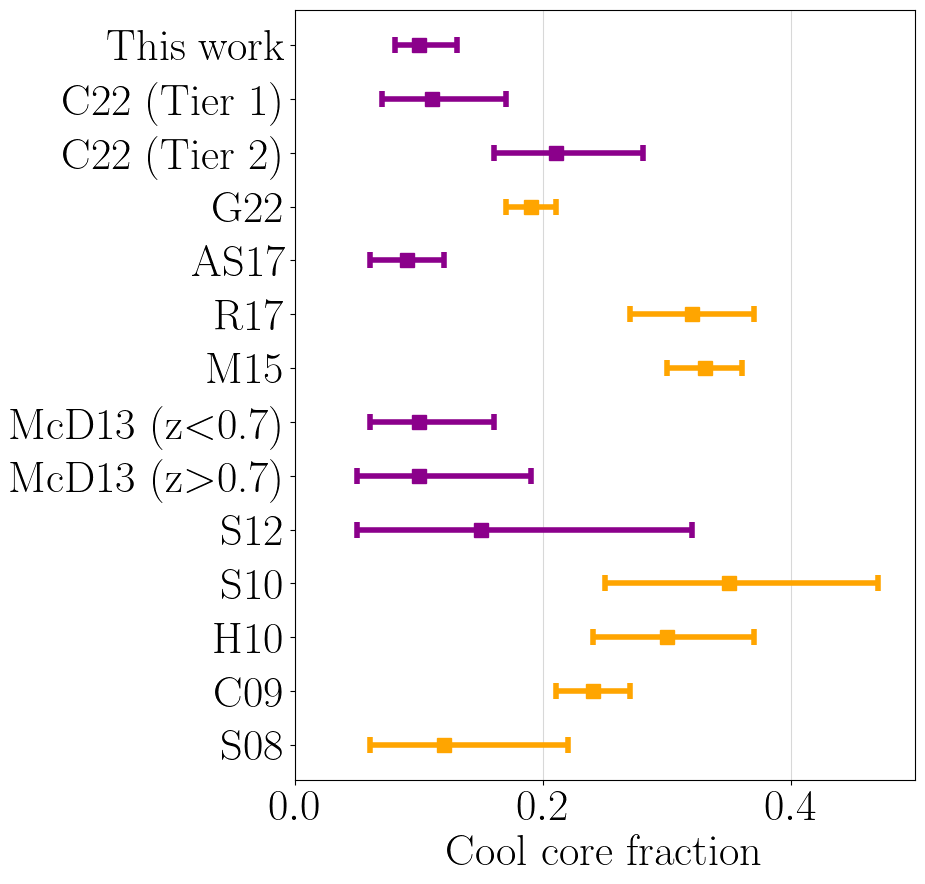}
\caption{A comparison of the cool core fraction calculated using surface brightness concentration for the CEREAL sample compared to the literature. For this figure, we define cool core systems as those with strong cool cores (c$_\mathrm{SB,~phys}>0.155$), and all errorbars are 1$\sigma$ confidence intervals calculated using binomial probabilities. For CHEX-MATE, we convert the c$_\mathrm{SB,~phys}$ threshold to a c$_\mathrm{SB,~scaled}$ threshold using the best-fit line from Figure \ref{fig:csb}. For any samples which use cool core proxies other than c$_\mathrm{SB}$, such as central entropy, we use relations from \cite{2010A&A...513A..37H} to convert to c$_\mathrm{SB}$ and apply the same cut. The referenced samples are: CHEX-MATE Tier 1 and Tier 2 \citep[C22;][]{2022A&A...665A.117C}, eFEDS \citep[G22;][]{2022A&A...661A..12G}, AS17 \citep{2017ApJ...843...76A}, MACS \citep[R17;][]{2017MNRAS.468.1917R}, M15 \citep{2015MNRAS.449..199M}, McD13 \citep{2013ApJ...774...23M}, S12 \citep{2012ApJ...761..183S}, 400D \citep[S10;][]{2010A&A...521A..64S}, HIFLUGCS \citep[H10;][]{2010A&A...513A..37H}, ACCEPT \citep[C09;][]{2009ApJS..182...12C}, and S08 \citep{2008A&A...483...35S}. The point types are color-coded with orange corresponding to X-ray selection and purple corresponding to SZ selection. In general, the cool core fraction tends to agree between publications that use the same cluster selection.
\label{fig:ccf}}
\end{figure}

\begin{figure*}[htb]
\plotone{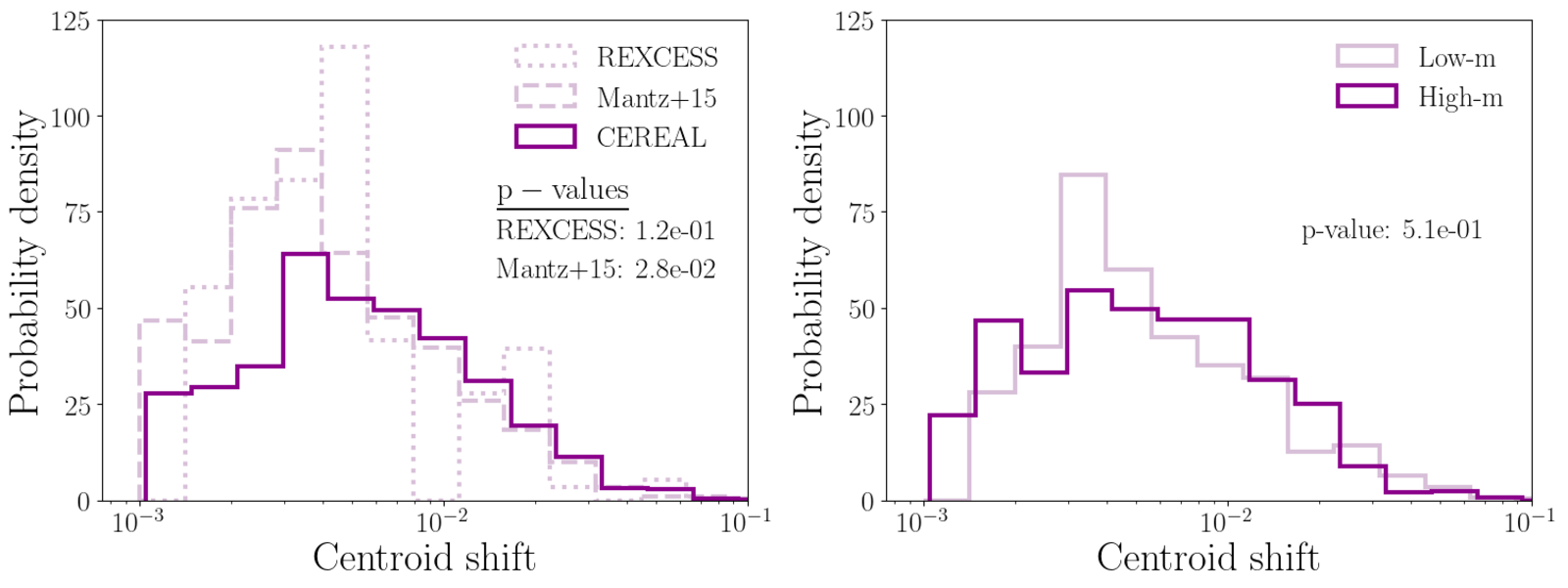}
\caption{Left: The distributions of centroid shift values for the CEREAL sample, the REXCESS sample \citep{2010A&A...514A..32B}, and the sample from \cite{2015MNRAS.449..199M}. Right: The distributions of centroid shifts for the CEREAL mass subsamples. Each plot shows the p-value for the null hypothesis that the measured distributions are drawn from the same underlying distribution, calculated using a K-S test. There is a statistically significant difference between the CEREAL and Mantz+15 samples, but there is no significant difference between the CEREAL and REXCESS clusters or} the high-mass and low-mass CEREAL clusters.
\label{fig:cent}
\end{figure*}

\section{Results} \label{sec:results}

\subsection{Cool core fraction}

We first compare the different methods of calculating concentration values using physically-sized apertures versus ones that scale with radius in Figure \ref{fig:csb}. We see a rough correlation between the two values, though there is significant scatter, leading to different classifications for clusters. In particular, many clusters which would not be considered cool cores using the former method do surpass the equivalent threshold using the latter method. We will examine this further in  \S\ref{sec:discussion}. We also note a number of clusters (12 in total) with scaled concentrations falling outside of the 0-1 range. Since the outer aperture for this calculation extends to the full radius of the cluster, the background subtraction can cause the fluxes in this region to be very low or even over-corrected to negative values.

We use the surface brightness concentration values in order to estimate the cool core fraction both for the entire sample as well as for the high- and low-mass subsamples. The distributions are shown in Figure \ref{fig:conc}. Using designations from \citet{2008A&A...483...35S} of $0.075 < $c$_\mathrm{SB,~phys}<0.155$ for moderate cool cores and c$_\mathrm{SB,~phys}>0.155$ for strong cool cores, we find 22 moderate cool cores and 10 strong cool cores in the high-mass sample of 108 clusters. We also find 16 moderate cool cores and 7 strong cool cores in the low-mass sample of 61 clusters. The inferred fractions are summarized in Table \ref{tab:coolcore}. Over the full sample, we find a cool core fraction (including moderate and strong cool cores) of $0.33^{+0.04}_{-0.04}$. In the left panel of Figure \ref{fig:conc}, we compare the distribution of concentration values for CEREAL clusters to those measured in various X-ray-selected samples, including HIFLUGCS \citep{2010A&A...513A..37H}, MACS \citep{2017MNRAS.468.1917R}, and the 400 SD survey \citep{2007ApJS..172..561B, 2010A&A...521A..64S}. In the right panel, we compare the CEREAL subsamples to the CHEX-MATE subsamples \citep{2022A&A...665A.117C}. Based on the Kolmogorov-Smirnov (K-S) test, we find that the distribution of concentrations in the CEREAL sample is inconsistent with those found for most X-ray-selected samples at high significance ($p < 0.05$). There is evidence, however, that the CEREAL distribution is consistent with the 400 SD distribution, which makes sense, given that the selection of the 400 SD sample spans a wider mass range. A similar agreement between CHEX-MATE and 400 SD was observed in \cite{2022A&A...665A.117C}. Note that the X-ray-selected samples are all consistent with each other. The CEREAL high-m distribution is consistent with the high-mass CHEX-MATE Tier 2 subsample, but there is a discrepancy between the CEREAL high-m clusters and the low-mass CHEX-MATE Tier 1 clusters.

In Table \ref{tab:coolcore} and Figure \ref{fig:ccf}, we summarize the cool core fractions derived from a variety of samples in the literature, both X-ray- and SZ-selected. In general, SZ-selected samples have cool core fractions of $\sim$25\% and strong cool core fractions of $\sim$10\%, while X-ray-selected samples have cool core fractions of $\sim$50\% and strong cool core fractions of $\sim$25\%. We emphasize again that, in this paper, we define a cool core cluster as a cluster with a high X-ray concentration. \cite{2010A&A...513A..37H} show the methodology used (e.g., central entropy, surface brightness concentration, etc) can have a strong effect on the derived cool core fraction. This table demonstrates broad consistency between works, provided the samples are selected in similar ways, which is reassuring. We conclude that the CEREAL sample, which is based on complete X-ray follow-up of an SZ-selected sample, contains significantly more low-concentration systems than the aforementioned X-ray-selected samples. We will discuss these results further in \S5.

Finally, we find no evidence that the distribution of X-ray concentrations is different between the low-mass and high-mass subsamples (Figure \ref{fig:conc}; right panel). Based on the K-S test, we find a $p$ value of 0.68, implying that the two subsamples are consistent with being drawn from the same parent distribution. This result indicates that, over the mass range covered here, the cool core fraction is not a strong function of mass.

\subsection{Dynamical state}

As with the concentration values, we use the distributions of centroid shifts ($w$) to understand differences between populations (see Figure \ref{fig:cent}). We compare the entire CEREAL sample to the clusters in the X-ray-selected Representative \textit{XMM-Newton} Cluster Structure Survey (REXCESS) sample \citep{2010A&A...514A..32B} and to the 361 clusters in the sample from \cite{2015MNRAS.449..199M}. Using the K-S test to test the null hypothesis that the CEREAL clusters are drawn from the same parent distribution as these X-ray-selected samples, we find a small ($< 5\%$) $p$ value when comparing to the sample from \cite{2015MNRAS.449..199M}, but there is some agreement between the CEREAL and REXCESS samples.

\begin{figure}[htb]
\plotone{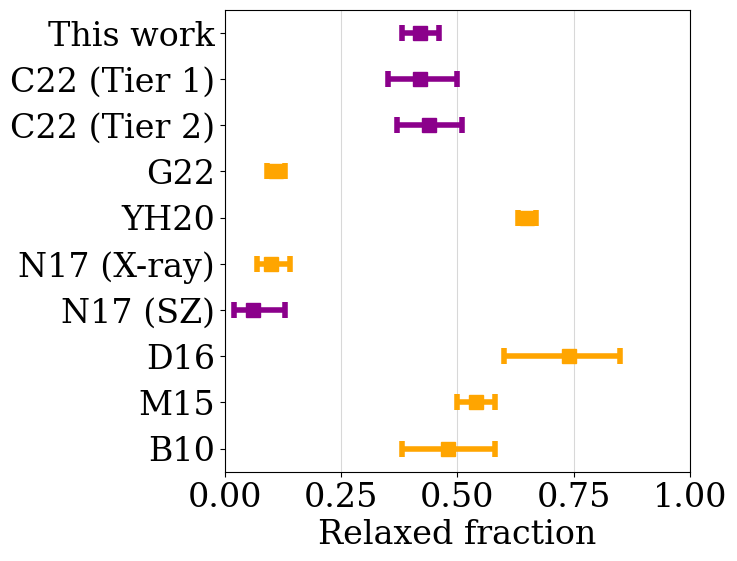}
\caption{A comparison of the relaxed fraction calculated using centroid shift for the CEREAL sample as well as various literature samples. For this figure, we define relaxed systems as those with $w<0.01$, and all errorbars are $1\sigma$ confidence intervals calculated using binomial probabilities to reflect sample size errors. The referenced samples are: CHEX-MATE Tier 1 and Tier 2 \citep[C22;][]{2022A&A...665A.117C}, the eFEDS sample \citep[G22;][]{2022A&A...661A..12G}, that from \citeauthor{2020MNRAS.497.5485Y} (YH20; \citeyear{2020MNRAS.497.5485Y}), that from \citeauthor{2017ApJ...841....5N} (N17; \citeyear{2017ApJ...841....5N}), that from \citeauthor{2016ApJ...819...36D} (D16; \citeyear{2016ApJ...819...36D}), that from \citeauthor{2015MNRAS.449..199M} (M15; \citeyear{2015MNRAS.449..199M}), and the REXCESS sample \citep[B10;][]{2010A&A...514A..32B}. The samples are color-coded with orange corresponding to X-ray selection and purple corresponding to SZ selection.
\label{fig:dynstate}}
\end{figure}

Using the centroid shift cut of $w=0.01$ from \cite{2010A&A...514A..32B}, we find that 46 of the high-mass CEREAL clusters and 25 of the low-mass CEREAL clusters are classified as relaxed, with an overall relaxed fraction of $0.42^{+0.04}_{-0.04}$. These fractions, and a comparison to the literature, are summarized in Table \ref{tab:dynstate} and Figure \ref{fig:dynstate}. This figure demonstrates that the relaxed fraction is much less consistent between publications, perhaps due to the wide variety of choices that are made in calculating the centroid shift compared to the concentration, such as differences in centering, estimation of cluster radii, choice of apertures, background handling, and handling of missing data. The measurement errors on the centroid shift values, which in some cases are quite large, are not included in these comparisons because we do not have these data available for all referenced samples; taking these errors into consideration may help resolve some of the tension between the different measurements of the relaxed fraction.

Consistent with the concentration, we find no statistically significant difference between the low-mass and high-mass CEREAL subsamples with regards to their centroid shift distributions. This implies that, over the mass range covered here, the observed dynamical states of clusters do not depend strongly on their total mass.

\subsection{Central point sources}

In \cite{2025ApJ...988...24W}, we presented the discovery of an X-ray-bright nucleus in the central galaxy of Abell 1885. Based on preliminary analysis of the CEREAL sample, we estimated the fraction of clusters hosting X-ray-bright central AGN with L$_\mathrm{2-10~keV}>10^{42}$ erg/s to be no more than 4.1\%. Here, we present a more uniform analysis of the central X-ray properties in the complete CEREAL sample, considering the excess luminosity of the nuclear regions of the clusters.

\begin{figure*}[htb]
\includegraphics[width=0.99\textwidth]{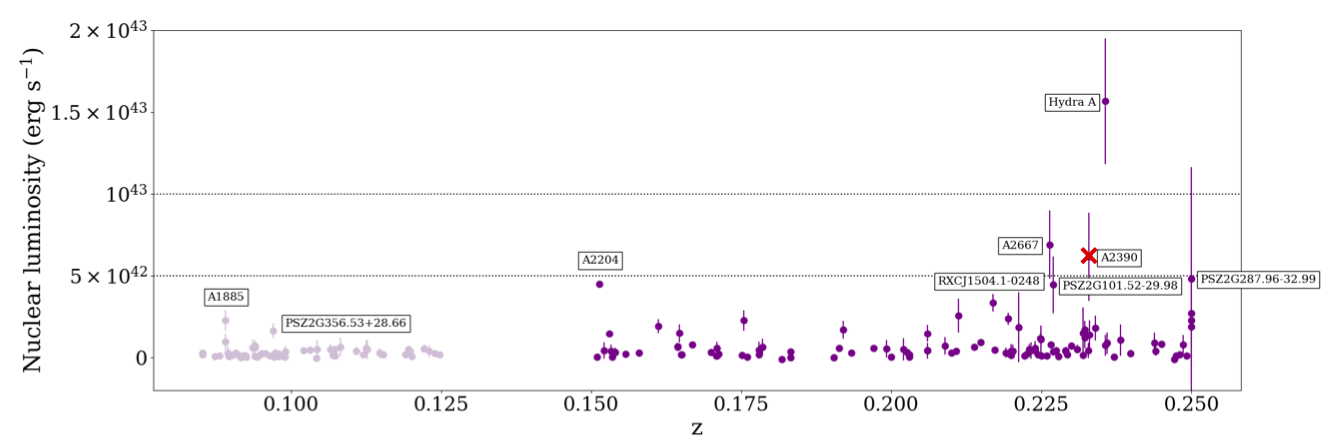}
\caption{Nuclear luminosity in the 2-10 keV band (based on excess flux in a 1 arcsec region centered on the cluster peak) as a function of redshift for the CEREAL clusters. The low-mass sample is shown in the lighter color, the high-mass sample is shown in the darker color, and the brightest central point source systems in each subsample have been labeled. Thresholds of interest at $5\times10^{42}$ and $10^{43}$ erg/s have been shown for references. A2390 is vetoed as an AGN detection because the X-ray peak is more than 1'' offset from the BCG.}
\label{fig:nuc}
\end{figure*}

Unlike with centroid shift and concentration, the excess X-ray luminosity often has uncertainties that are similar in magnitude to the population scatter. This is because some of the systems have very small numbers of excess counts in the core, driving up the error from counting statistics. These uncertainties need to be accounted for in addition to the binomial uncertainty which arises from dividing the continuous distribution of nuclear luminosities into two discrete populations of ``detections" and ``non-detections" in order to accurately capture the uncertainty in the fraction of clusters hosting a central AGN. We bootstrap over the measured X-ray luminosity uncertainties and determine the binomial distribution for the number of X-ray-bright nuclei for each bootstrap. We repeat this process 5000 times, combining these distributions into an average distribution that has appropriate contributions from measurement uncertainty and counting statistics. We report the expected number of successes for the resulting distribution as the true underlying fraction of clusters with nuclear luminosities above that threshold, with errors quoted as the $1\sigma$ confidence interval. We test this error propagation procedure for a range of success fractions, measurement errors, and numbers of observations, and find that it behaves well (accurately recovering the success fraction within a $1\sigma$ confidence interval) in the regime in which we are operating: 100-200 observations with an average measurement error of $<100\%$. In Table \ref{tab:ps}, we present the observed number of detections as well as the inferred detection fractions of clusters with high nuclear luminosities for a variety of threshold values. We calculate these fractions for the mass subsamples as well as for the entire sample divided into cool-core and non-cool core clusters based on $c_\mathrm{SB}$.

The nuclear luminosities of all of the systems are shown in Figure \ref{fig:nuc}. In the low-mass sample, the most outstanding systems are A1885, which we previously discovered, and PSZ2G356.53+28.66. The latter is a candidate for further follow-up in search of evidence of AGN feedback. In the high-mass sample, the systems with the highest luminosities are well-known systems which have been previously found to have evidence of strong AGN feedback, such as Hydra A, and A2667. However, even among these most luminous systems, we note that no cluster exceeds $L_\mathrm{nuc}>10^{44}\mathrm{~erg~s}^{-1}$, suggesting that supermassive black holes accreting near the Eddington limit may be exceptionally rare in low-z cluster environments. One critical limitation of this X-ray imaging analysis is that highly-accreting AGN are often subject to substantial obscuration in the X-ray by dusty torii, which could cause us to underestimate the occurrence of intrinsically bright AGN. A spectroscopic X-ray analysis could help correct for this, but ultimately, multiwavelength observations would provide the most robust constraint on the central AGN fraction in clusters.

\begin{figure}[tb]
\centering
\includegraphics[width=0.99\linewidth]{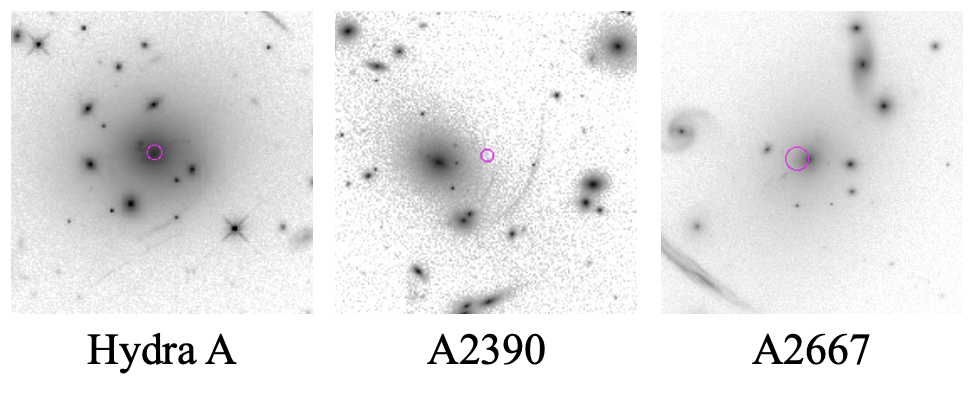}
\caption{\textit{Hubble} images of the 3 CEREAL clusters with $L_\mathrm{nuc}>5\times 10^{42}\mathrm{~erg~s}^{-1}$: Hydra A (F140W), A2390 (F850LP), and A2667 (F850LP). The circles represent 1$^{\prime\prime}$ regions centered on the X-ray peaks. In the case of A2390, the peak is more than 1$^{\prime\prime}$ offset from the BCG, so this is vetoed as a central AGN system.
\label{fig:AGNimages}}
\end{figure}

To verify that our analysis is capturing X-ray-luminous AGN in the central cluster galaxy, we present \textit{Hubble} images of the 3 clusters with $L_\mathrm{nuc}>5\times 10^{42}\mathrm{~erg~s}^{-1}$ in Figure \ref{fig:AGNimages}. The X-ray peaks used for the nuclear luminosity calculation are indicated on each image. We veto any clusters where the X-ray peak is more than 1'' offset from the BCG, which is the case for A2390.

\begin{figure}[tb]
\plotone{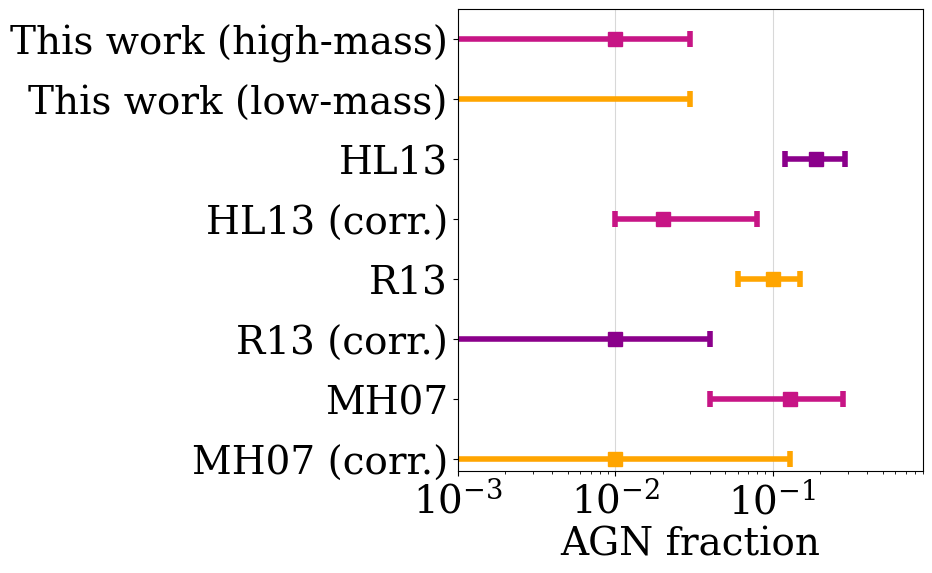}
\caption{A comparison of the X-ray-bright AGN fraction calculated for the CEREAL sample as well as various literature samples. For this figure, we define systems with central X-ray-bright AGN as those with $L_\mathrm{nuc,~2-10~keV}>10^{43} \mathrm{~erg~s}^{-1}$, and all errorbars are $1\sigma$ confidence intervals calculated using binomial probabilities to reflect sample size errors. The reference samples are that from \citeauthor{2013MNRAS.431.1638H} (HL13; \citeyear{2013MNRAS.431.1638H}), \citeauthor{2013MNRAS.432..530R} (R13; \citeyear{2013MNRAS.432..530R}), and \citeauthor{2007MNRAS.381..589M} (MH07; \citeyear{2007MNRAS.381..589M}). Note that these samples are feedback-selected and therefore highly biased towards cool core systems, which tend to host central AGN. To correct for this, we multiply the literature values by the strong cool core fraction estimated in this work, which brings them in line with our value.
\label{fig:agnfrac}}.
\end{figure}

We compare our AGN fraction to various literature values in Table \ref{tab:agnfrac} and Figure \ref{fig:agnfrac}. In general, the occurrence rate of X-ray-bright AGN at the centers of clusters is not well-studied, and this represents the first attempt to quantify it in a well-selected sample. Other works \citep[e.g.,][]{2013MNRAS.431.1638H} find a significantly higher fraction of X-ray-bright central AGN in samples of clusters selected to have strong feedback. When these studies are corrected to include non-cool core clusters, we find overall good agreement that rapidly-accreting central SMBHs are found in only $\sim$1\% of clusters.

\section{Discussion} \label{sec:discussion}

In this work, we find that the majority of SZ-selected clusters have low central X-ray surface brightness, indicating that they are moderate or non-cool cores. This is consistent with previous studies, which find X-ray-selected samples to have a significantly higher fraction of cool core clusters than SZ-selected samples \citep[e.g.,][]{2011A&A...526A..79E, 2017MNRAS.468.1917R, 2017ApJ...843...76A}. We find a continuous distribution of concentration values from moderate cool core values all the way up to strong values, disfavoring a CC/NCC bimodality. We emphasize that the use of surface brightness concentration as a proxy for identifying cool cores is useful in this analysis to compare to the literature, but it is likely not the ideal measurement to use in future evolutionary studies. In particular, we see that the classifications of clusters as strong, moderate, or non cool cores is subject to the choice of apertures used. This is even more relevant when extending this work to higher redshifts, where the physical sizes of the clusters become much smaller. Follow-up work using the thermodynamic profiles (White et al. in prep) will further assess the cool core fraction based on central entropy and cooling time and look for bimodality in cluster cooling state.

Analyzing the cool core fraction as a function of redshift gives us a peek into the history of feedback from active galactic nuclei (AGN). In particular, we are interested in the balance between cooling of the intracluster medium (ICM) and mechanical energy output from the AGN of the brightest cluster galaxy (BCG). The redshift dependence of this relationship could potentially reveal the onset of AGN feedback early in the history of the universe. Figure 10 from \citet{2021ApJ...918...43R} shows no significant evolution in the cool core fraction, with a value of $43_{-6}^{+7}$\% for clusters at $0.25 < z < 1.3$. Our measurement at z$\sim$0.15, which is $30_{-5}^{+5}$\% for the high-mass sample (which would be the descendants of the clusters studied by \cite{2021ApJ...918...43R}), including both strong and moderate cool cores, is in weak tension with this measurement. However, this earlier work used central density ($n_{e,0} > 0.015$ cm$^{-3}$) as a cool core indicator; if we apply the moderate cool core definition from \cite{2010A&A...513A..37H}, $n_{e,0} > 0.05$ cm$^{-3}$, the high-$z$ cool core fraction from \cite{2021ApJ...918...43R} would become $8_{-4}^{+12}$\%, highlighting the importance of consistent methodology across samples. We note that central density correlates with mass, which must be taken into consideration if using it for comparative studies between samples spanning different mass ranges. In this case, the CEREAL clusters were selected in part to facilitate a comparison with the sample from \cite{2021ApJ...918...43R}. Moving forward, the large, well-defined sample of clusters presented here will facilitate anchoring of future analyses of the evolution of cool cores (McDonald et al.\ in prep).

Recent work by \cite{2025A&A...694A.232G} comparing observations to simulations found a lower fraction of cool cores among massive clusters when compared to low-mass clusters and galaxy groups. Matching our samples in mass, \cite{2025A&A...694A.232G} find a cool core fraction of $\sim$0.35 for low-mass clusters (M$_{500}$ = 2.5--4 $\times$ 10$^{14}$ M$_{\odot}$) and $\sim$0.2 for high-mass clusters (M$_{500}$ = 4--10 $\times$ 10$^{14}$ M$_{\odot}$). While the low-mass fraction agrees well with the CEREAL sample (0.38$_{-0.07}^{+0.07}$), their high-mass fraction falls outside of our 1$\sigma$ confidence interval (0.30$_{-0.05}^{+0.05}$). We note that this work by \cite{2025A&A...694A.232G} uses core temperature rather than concentration to delineate cool cores, so a more direct comparison will be forthcoming alongside our thermodynamical analysis. Over the mass range covered by the CEREAL sample, \cite{2025A&A...694A.232G} predict a factor of 4 change in the cool core fraction, which is inconsistent with what we find based on concentration. Instead, we find no evidence for a mass dependence of the cool core fraction, suggesting that whatever process drives the formation/destruction of cool cores is not strongly mass dependent.

We note that the Malmquist bias arising from the flux-limited selection of objects into the X-ray sample can play a significant role in the observed discrepancy of the cool core fractions. This effect was considered in \cite{2017ApJ...843...76A}. There is a systematic difference in the X-ray luminosity between cool core and non-cool core clusters of the same mass. \cite{2017ApJ...843...76A} estimate that the cool core clusters are, on average, $\sim$ 70\% more luminous and hence have a greater probability to pass selection into an X-ray flux limited sample. They further estimate that the Malmquist bias can boost the number of cool core clusters in a sample by a factor of $\sim$2, thus potentially explaining most of the differences we observe in this work. Note that most of the cosmological analyses of the cluster mass function based on X-ray samples explicitly include Malmquist bias in their statistical models \citep[e.g.,][]{2009ApJ...692.1033V, 2009ApJ...692.1060V, 2010MNRAS.406.1759M, 2010MNRAS.406.1773M} Therefore, the morphological differences we observe do not necessarily indicate a problem with these analyses.

We see no significant difference in the number of morphologically disturbed/asymmetric systems between the low-mass and high-mass CEREAL subsamples. Dark matter only simulations predict that the merger rate should be only a weak function of mass \citep[$\propto M^{0.133}$;][]{2010MNRAS.406.2267F}. Such a weak mass dependence, coupled with the factor-of-two increase in average mass between our two samples leads to a prediction of 10\% more mergers in the high-mass bin compared to the low-mass bin. Given that these samples have an average of $\sim$80 clusters, and $\sim$60\% are disturbed, we would predict an extra $\sim$4 merging clusters in the high-mass sample compared to the low-mass sample. This is significantly smaller than the Poisson uncertainty. Thus, the consistency that we observe in the relaxed fraction between the high-mass and low-mass CEREAL subsamples is fully consistent with $\Lambda$CDM predictions. Further, the lack of mass dependence for the dynamical state is consistent with previous observations of SZ- and X-ray-selected samples \citep[e.g.,][]{2010A&A...514A..32B, 2017ApJ...846...51L, 2017MNRAS.468.1917R, 2019A&A...628A..86B}. This will be an important point for upcoming evolutionary studies which compare high-redshift cluster samples to our low-redshift anchor. Because clusters undergo hierarchical structure formation, they become more massive over time. However, given that statistics like the cool core fraction and merger rate do not appear to trend with mass, any changes that we observe in these will not be due to the growth in mass but rather due to true cosmic evolution. 

In terms of central point sources, we find a significant increase in the nuclear X-ray luminosity with cluster mass (Figure \ref{fig:nuc}). Following \cite{2018ApJ...858...45M}, we expect the central black hole mass to scale with halo mass like $M_\mathrm{BH} \propto M_{500}^{0.32}$. Given the factor-of-two difference in masses between the high-mass and low-mass sample, we expect a $\sim$25\% increase in the Eddington rate for the black holes in the centers of these systems. We observe a significantly larger increase, with the high-mass subsample spanning an order of magnitude larger range in nuclear luminosity than the low-mass sample. This may imply that central AGN in more high-mass clusters are accreting gas more rapidly, perhaps due to more efficient cooling in these systems or a higher rate of mergers. We compare to the population of clusters from \cite{2013MNRAS.431.1638H}, which was selected on the presence of strong AGN feedback. The nuclear luminosities that they measured for these extreme systems ranged from $10^{41}$ to $10^{45}$ erg s$^{-1}$, with most having L$_\mathrm{nuc} > 10^{42}$ erg s$^{-1}$. We find a total of 33 systems across both CEREAL mass subsamples with such high nuclear luminosities. Given that this earlier work was based on a sample of clusters already identified to have strong AGN feedback, it is unsurprising that they measure such a high fraction of clusters with central X-ray-bright point sources. When we calculate the AGN fractions for cool core systems compared to non-cool core systems, we see evidence that cool cores preferentially host X-ray-bright central AGN, though we do note that central AGN could contaminate the $c_\mathrm{SB}$ values used here to identify cool cores, which could contribute to this correlation. Therefore, if we assume that the selection in \cite{2013MNRAS.431.1638H} is effectively a strong cool core selection, we can correct their sample to include the non-cool cores, finding broad agreement (see Figure \ref{fig:agnfrac}). Based on the CEREAL sample, we estimate the occurrence of Perseus-like systems with low redshift, high mass, and L$_\mathrm{nuc} > 10^{43}$ erg s$^{-1}$ to be $\sim 0.7\%$, and we can place an upper limit on even brighter systems like Cygnus A, 3C 295, and H1821+643 with L$_\mathrm{nuc} > 10^{44}$ erg s$^{-1}$ at no more than 1.6\% of massive clusters at low redshift at $1\sigma$ confidence.

\section{Conclusion} \label{sec:conclusion}

We present, for the first time, the CEREAL sample: a large, well-understood, SZ-selected sample of galaxy clusters observed in the X-ray with \textit{Chandra} which spans nearly an order of magnitude in mass and a narrow redshift range around z$\sim$0.15. This sample, with its clean selection and uniform follow-up, is ideally suited to studying the demographics of galaxy clusters at high redshifts, which are typically identified using SZ selection as well. In this work, we focus on an X-ray surface brightness analysis of these data, finding:

\begin{itemize}
    \item The cool core fraction (c$_\mathrm{SB,~phys} > 0.075$) and strong cool core fraction (c$_\mathrm{SB,~phys}>0.155$) in the CEREAL sample, measured using surface brightness concentration, are $0.22_{-0.03}^{+0.04}$ and $0.10_{-0.02}^{+0.03}$, respectively, which are significantly lower than what is found in X-ray-selected samples and consistent with findings from other SZ-selected samples. 
    
    \item Over the mass range covered by the CEREAL sample ($2-10\times 10^{14}$ M$_{\odot}$), we find no significant dependence of the cool core fraction on mass.
    
    \item The merger/disturbed rate, inferred from the distribution of centroid shifts, is $0.58_{-0.04}^{+0.04}$. We find that there is considerably more scatter in published values of the merger rate, likely due to the higher complexity of this measurement compared to the surface brightness concentration.
    
    \item Over the mass range covered by the CEREAL sample ($2-10\times 10^{14}$ M$_{\odot}$), we find no significant dependence of the merger/disturbed fraction on mass. This is consistent with $\Lambda$CDM simulations, which predict only a weak mass dependence of the merger rate and would require a much larger sample to detect, as well as previous observations.
    
    \item Based on excess central luminosity, the black hole accretion rate in the central cluster galaxy is notably higher in more massive clusters. This effect is more than can be explained only by higher black hole masses, pointing to increased accretion efficiency in these systems.
    
    \item Clusters with X-ray-bright central point sources are rare. In 169 observations, only one such system has been newly identified. We place an upper limit of 1.6\% on the occurrence of extremely luminous central AGN ($L_{\mathrm{nuc,2-10~keV}} > 10^{44}$ erg/s) in massive clusters at $z\sim0$. Our observations are consistent with literature values when correcting previously-published works for selection bias.
    
\end{itemize}

\noindent Future work on the CEREAL sample (White et al. in prep) will focus on thermodynamic profiles, which will facilitate a more accurate cool core fraction based on central entropy, probe the burstiness of AGN feedback using the scatter in entropy/density profiles, and constrain the universal pressure and temperature profiles for low-redshift galaxy clusters. This work will further solidify the use of the CEREAL sample as an anchor for evolutionary studies.

\begin{acknowledgments}
Support for this work was provided to L.W.\ and M.M.\ by NASA through Chandra Award Numbers GO2-23117X and GO4-25083 issued by Chandra, which is operated by the Smithsonian Astrophysical Observatory for and on behalf of the National Aeronautics Space Administration under contract NAS803060.
\end{acknowledgments}

\vspace{5mm}
\facilities{CXO, Planck, HST}

\software{astropy \citep{2013A&A...558A..33A, 2018AJ....156..123A, 2022ApJ...935..167A},
          CIAO \citep{2006SPIE.6270E..1VF}
          }

\bibliography{references}{}
\bibliographystyle{aasjournal}

\begin{appendix}

\setcounter{table}{0}
\renewcommand{\thetable}{A\arabic{table}}

\section{Cluster Demographics Tables}
Here we summarize the various demographics of CEREAL clusters, including the cool core fraction, the relaxed fraction, and the AGN-hosting fraction, compared to those from the literature, as presented in the paper. These tables have been moved to the Appendix for clarity.

\setlength{\tabcolsep}{12pt}
    \begin{table*}[htb]
    \centering
    \begin{tabular}{c|c|c c c}
    \hline
         \textbf{Sample} & \textbf{Selection} & \textbf{f}$_\mathrm{\textbf{\tiny NCC}}$ & \textbf{f}$_\mathrm{\textbf{\tiny mod}}$ & \textbf{f}$_\mathrm{\textbf{\tiny SCC}}$ \\ \hline
         Low-mass CEREAL & SZ & $0.62^{+0.07}_{-0.07}$ & $0.26^{+0.07}_{-0.06}$ & $0.11^{+0.06}_{-0.04}$ \\
         High-mass CEREAL & SZ & $0.70^{+0.05}_{-0.05}$ & $0.20^{+0.05}_{-0.04}$ & $0.09^{+0.04}_{-0.03}$ \\
         Overall CEREAL & SZ & $0.67^{+0.04}_{-0.04}$ & $0.22^{+0.04}_{-0.03}$ & $0.10^{+0.03}_{-0.02}$ \\
         & & & & \\

         Campitiello et al. 2022 (Tier 1) & SZ &
         $0.56^{+0.07}_{-0.08}$ &
         $0.33^{+0.08}_{-0.07}$ &
         $0.11^{+0.06}_{-0.04}$ \\
         Campitiello et al. 2022 (Tier 2) & SZ &
         $0.52^{+0.07}_{-0.07}$ &
         $0.26^{+0.07}_{-0.06}$ &
         $0.21^{+0.07}_{-0.05}$ \\
         Andrade-Santos et al. 2017 & SZ & $0.69^{+0.05}_{-0.05}$ &  $0.23^{+0.05}_{-0.04}$& $0.09^{+0.03}_{-0.03}$ \\
         McDonald et al. 2013 (z$<$0.7) & SZ & $0.74^{+0.07}_{-0.08}$ & $0.16^{+0.07}_{-0.05}$ & $0.10^{+0.06}_{-0.04}$ \\
         McDonald et al. 2013 (z$>$0.7) & SZ & $0.65^{+0.09}_{-0.11}$ & $0.26^{+0.10}_{-0.08}$ & $0.10^{+0.09}_{-0.05}$ \\
         Semler et al. 2012 & SZ & $0.54^{+0.16}_{-0.17}$ & $0.31^{+0.18}_{-0.14}$ & $0.15^{+0.17}_{-0.10}$ \\
& & & & \\
         Ghirardini et al. 2022 & X-ray & $0.51^{+0.03}_{-0.03}$ & $0.30^{+0.03}_{-0.03}$ & $0.19^{+0.02}_{-0.02}$ \\
         Rossetti et al. 2017 & X-ray & $0.45^{+0.05}_{-0.05}$ & $0.23^{+0.05}_{-0.04}$ & $0.32^{+0.05}_{-0.05}$ \\
         Mantz et al. 2015 & X-ray & $0.39^{+0.03}_{-0.03}$ & $0.29^{+0.03}_{-0.03}$ & $0.33^{+0.03}_{-0.03}$ \\
         Santos et al. 2010 & X-ray & $0.42^{+0.12}_{-0.11}$ & $0.23^{+0.11}_{-0.09}$ & $0.35^{+0.12}_{-0.10}$ \\
         Hudson et al. 2010 & X-ray & $0.45^{+0.07}_{-0.07}$ & $0.25^{+0.07}_{-0.06}$ & $0.30^{+0.07}_{-0.06}$ \\
         Cavagnolo et al. 2009 & X-ray & $0.45^{+0.04}_{-0.03}$ & $0.31^{+0.03}_{-0.03}$ & $0.24^{+0.03}_{-0.03}$ \\
         Santos et al. 2008 & X-ray & $0.35^{+0.12}_{-0.10}$ & $0.54^{+0.11}_{-0.12}$ & $0.12^{+0.10}_{-0.06}$ \\
         \hline
    \end{tabular}
    \caption{Cool core fractions for the mass subsamples, the overall CEREAL sample, and a number of literature samples referenced in Figure \ref{fig:ccf}. Note that literature proxies for cool-coredness have been converted to concentration (c$_\mathrm{\tiny SB,~phys}$) using scaling relations from \cite{2010A&A...513A..37H}. We use the designations from \citet{2008A&A...483...35S} for non-cool cores (NCC), moderate cool cores (mod), and strong cool cores (SCC). The errors are 1$\sigma$ confidence intervals reflecting the binomial statistics based on sample size.}
    \label{tab:coolcore}
\end{table*}

\begin{table*}[htb]
    \centering
    \begin{tabular}{c|c|c c}
    \hline
         \textbf{Sample} & \textbf{Selection} & \textbf{f}$_\mathrm{\textbf{\tiny relaxed}}$ & \textbf{f}$_\mathrm{\textbf{\tiny disturbed}}$ \\ \hline
         Low-mass CEREAL & SZ & $0.41^{+0.07}_{-0.07}$ & $0.59^{+0.07}_{-0.07}$ \\
         High-mass CEREAL & SZ & $0.43^{+0.05}_{-0.05}$ & $0.57^{+0.05}_{-0.05}$ \\
         Overall CEREAL & SZ & $0.42^{+0.04}_{-0.04}$ & $0.58^{+0.04}_{-0.04}$ \\ 
         & & &\\
         Campitiello et al. 2022 (Tier 1) & SZ &
         $0.42^{+0.08}_{-0.07}$ &
         $0.58^{+0.07}_{-0.08}$ \\
         Campitiello et al. 2022 (Tier 2) & SZ &
         $0.44^{+0.07}_{-0.07}$ &
         $0.56^{+0.07}_{-0.07}$ \\
         Nurgaliev et al. 2017 (SZ) & SZ & $0.10^{+0.04}_{-0.03}$ &  $0.90^{+0.03}_{-0.04}$ \\
         & & & \\
         Ghirardini et al. 2022 & X-ray & $0.11^{+0.02}_{-0.02}$ & $0.89^{+0.02}_{-0.02}$ \\
         Yuan and Han 2020 & X-ray & $0.65^{+0.02}_{-0.02}$ & $0.35^{+0.02}_{-0.02}$ \\
         Nurgaliev et al. 2017 (X-ray) & X-ray & $0.06^{+0.07}_{-0.04}$ & $0.94^{+0.04}_{-0.07}$\\
         Donahue et al. 2016 & X-ray & $0.74^{+0.11}_{-0.14}$ & $0.26^{+0.14}_{-0.11}$ \\
         Mantz et al. 2015 & X-ray & $0.54^{+0.04}_{-0.04}$ & $0.46^{+0.04}_{-0.04}$ \\
         Bohringer et al. 2010 & X-ray & $0.48^{+0.10}_{-0.10}$ & $0.52^{+0.10}_{-0.10}$ \\
         \hline
    \end{tabular}
    \caption{Dynamical state fractions for the mass subsamples, the overall CEREAL sample, and a number of literature samples referenced in Figure \ref{fig:dynstate}. Note that literature values have been calculated from available data using our own designations and error propagation. We use the cut of \textit{w} $<$ 0.01 from \cite{2010A&A...514A..32B} to define a relaxed cluster. The errors are 1$\sigma$ confidence intervals reflecting the binomial statistics based on sample size.}
    \label{tab:dynstate}
\end{table*}

\begin{table*}[htb]
    \centering
    \begin{tabular}{c|c c|c c|c c|c c}
         \textbf{Nuclear Luminosity Threshold} & \multicolumn{2}{c|}{\textbf{Low-mass}} & \multicolumn{2}{c}{\textbf{High-mass}} &
         \multicolumn{2}{|c|}{\textbf{Cool cores}} &
         \multicolumn{2}{c}{\textbf{Non-cool cores}}\\ 
         \textbf{[erg/s]} & \textbf{N} & \textbf{\%} & \textbf{N} & \textbf{\%} & \textbf{N} & \textbf{\%} & \textbf{N} & \textbf{\%}\\
         \hline\hline
         $5 \times 10^{42}$ & 0 & $1.1^{+1.8}_{-0.8}$ & 2 & $1.6^{+1.5}_{-0.9}$ & 1 & $11.5^{+10.0}_{-6.6}$ & 1 & $0.5^{+0.8}_{-0.3}$ \\ \hline
         $10^{43}$& 0 & $1.1^{+1.8}_{-0.8}$ & 1 & $0.7^{+1.2}_{-0.5}$ & 1 & $8.0^{+9.9}_{-5.7}$ & 0 & $0.4^{+0.7}_{-0.3}$ \\ \hline
         $5 \times 10^{43}$& 0 & $1.1^{+1.8}_{-0.8}$ & 0 & $0.6^{+1.0}_{-0.5}$ & 0 & $4.8^{+7.5}_{-3.6}$ & 0 & $0.4^{+0.7}_{-0.3}$ \\ \hline
         $10^{44}$ & 0 & $1.1^{+1.8}_{-0.8}$ & 0 & $0.6^{+1.0}_{-0.5}$ & 0 & $4.8^{+7.5}_{-3.6}$ & 0 & $0.4^{+0.7}_{-0.3}$ \\
    \end{tabular}
    \caption{The observed number and inferred fraction of nuclear (1'') regions with excess flux above a specified luminosity threshold for the low-mass and high-mass CEREAL samples. We also report these fractions for the full sample divided into cool cores (13 total) and non-cool cores (156 total), using $c_\mathrm{SB}>0.155$ to define cool cores. Quoted errors are 1$\sigma$ confidence intervals reflecting uncertainty from both measurement error as well as binomial statistics.}
    \label{tab:ps}
\end{table*}

\begin{table*}[htb]
    \centering
    \begin{tabular}{c|c}
    \hline
         \textbf{Sample} & \textbf{f}$_\mathrm{\textbf{\tiny AGN}}$ \\ \hline
         Low-mass CEREAL &  $0.00^{+0.03}_{-0.00}$ \\
         High-mass CEREAL &  $0.01^{+0.02}_{-0.01}$  \\ 
         \\
         Hlavacek-Larrondo et al. 2013 & $0.19^{+0.10}_{-0.07}$ \\
         Russell et al. 2013 & $0.10^{+0.05}_{-0.04}$ \\
         Merloni and Heinz 2007 & $0.13^{+0.15}_{-0.09}$ \\
\\
         Hlavacek-Larrondo et al. 2013 (corrected) & $0.02^{+0.06}_{-0.01}$ \\
         Russell et al. 2013 (corrected) & $0.01^{+0.03}_{-0.01}$ \\
         Merloni and Heinz 2007 (corrected) & $0.01^{+0.12}_{-0.01}$ \\
         \hline
    \end{tabular}
    \caption{Central AGN fractions for the CEREAL mass subsamples and a number of literature samples referenced in Figure \ref{fig:agnfrac}. Note that literature values have been calculated from available data using our own designations and error propagation. We use the cut of L$_\mathrm{nuc,2-10~keV}>10^{43} \mathrm{~erg~s}^{-1}$ to define an X-ray-bright central AGN. The errors are 1$\sigma$ confidence intervals reflecting the binomial statistics based on sample size. Note that the literature samples are feedback-selected and therefore highly biased towards cool core systems. To correct for this, we multiply the literature values by the strong cool core fraction estimated in this work, which brings them in line with our value.}
    \label{tab:agnfrac}
\end{table*}

\clearpage

\section{Full CEREAL Sample and X-ray Surface Brightness Measurements}

\setcounter{table}{0}
\renewcommand{\thetable}{B\arabic{table}}

\setlength\LTleft{0pt}
\setlength\LTright{0pt}
\setlength{\LTcapwidth}{\textwidth}
\setlength{\tabcolsep}{3pt}
\begin{longtable}{@{\extracolsep{\fill}}c|c|c|c|c|c|c|c|c@{}}
\caption{Below, we summarize the full sample of CEREAL clusters, including their redshifts and masses from the Planck catalog \citep{2016A&A...594A..27P}, the location of the X-ray peak on the sky, and the measured concentration, centroid shift, and excess nuclear luminosity in the 2-10 keV band. The low-mass clusters are listed first, followed by the high-mass clusters. All errorbars represent 1$\sigma$ errors.} \\

& & \textbf{M}$_\mathrm{\textbf{\tiny 500}}$ & & & & & & \textbf{L}$_\mathrm{\textbf{\tiny nuc, 2-10~keV}}$ \\ 
    \textbf{Name} & \textbf{z} & [\textbf{10}$^{\mathrm{\textbf{\tiny 14}}}$ \textbf{M}$_\odot$] & \textbf{RA} & \textbf{dec} & \textbf{c}$_\mathrm{\textbf{\tiny SB,phys}}$ & \textbf{c}$_\mathrm{\textbf{\tiny SB,scaled}}$ & \textbf{w} & \textbf{[erg/s]} \\ \hline \hline
\endfirsthead
     & & \textbf{M}$_\mathrm{\textbf{\tiny 500}}$ & & & & & & \textbf{L}$_\mathrm{\textbf{\tiny nuc, 2-10~keV}}$ \\ 
    \textbf{Name} & \textbf{z} & [\textbf{10}$^{\mathrm{\textbf{\tiny 14}}}$ \textbf{M}$_\odot$] & \textbf{RA} & \textbf{dec} & \textbf{c}$_\mathrm{\textbf{\tiny SB,phys}}$ & \textbf{c}$_\mathrm{\textbf{\tiny SB,scaled}}$ & \textbf{w} & \textbf{[erg/s]} \\ \hline \hline
    \endhead

PSZ2G000.04+45.13 & 0.12 & 3.96 & 229.19 & -1.02 & 0.045 $\pm$ 0.011 & -0.610 $\pm$ -1.012 & 0.018 $\pm$ 0.006 & $3.8 \pm 3.2  \times 10^{41}$ \\
PSZ2G006.68-35.55 & 0.09 & 4.17 & 308.71 & -35.83 & 0.023 $\pm$ 0.003 & 0.366 $\pm$ 0.021 & 0.008 $\pm$ 0.009 & $2.8 \pm 4.0  \times 10^{41}$ \\
PSZ2G007.76-36.83 & 0.09 & 2.64 & 310.44 & -35.19 & 0.032 $\pm$ 0.010 & 0.215 $\pm$ 0.087 & 0.066 $\pm$ 0.009 & $1.6 \pm 2.0  \times 10^{41}$ \\
PSZ2G008.80-35.18 & 0.09 & 2.98 & 308.66 & -34.06 & 0.139 $\pm$ 0.012 & 0.648 $\pm$ 0.094 & 0.004 $\pm$ 0.010 & $5.9 \pm 5.1  \times 10^{41}$ \\
PSZ2G018.73+23.56 & 0.09 & 3.97 & 255.58 & -1.02 & 0.032 $\pm$ 0.005 & 0.299 $\pm$ 0.033 & 0.024 $\pm$ 0.011 & $9.6 \pm 5.0  \times 10^{41}$ \\
PSZ2G018.84+22.43 & 0.09 & 3.60 & 256.61 & -1.50 & 0.037 $\pm$ 0.005 & 0.444 $\pm$ 0.126 & 0.033 $\pm$ 0.011 & $1.3 \pm 1.5  \times 10^{41}$ \\
PSZ2G028.63+50.15 & 0.09 & 3.25 & 235.03 & 17.90 & 0.274 $\pm$ 0.019 & 0.851 $\pm$ 0.348 & 0.115 $\pm$ 0.007 & $-2.5 \pm 1.8  \times 10^{39}$ \\
PSZ2G033.46-48.43 & 0.09 & 4.08 & 328.08 & -19.61 & 0.138 $\pm$ 0.001 & 0.515 $\pm$ 0.013 & 0.058 $\pm$ 0.010 & $3.4 \pm 0.5  \times 10^{40}$ \\
PSZ2G037.05-70.55 & 0.09 & 2.22 & 351.52 & -24.12 & 0.262 $\pm$ 0.027 & 0.536 $\pm$ 0.157 & 0.012 $\pm$ 0.006 & $8.7 \pm 6.4  \times 10^{40}$ \\
PSZ2G041.43-66.81 & 0.11 & 2.71 & 348.11 & -21.55 & 0.091 $\pm$ 0.008 & 0.442 $\pm$ 0.058 & 0.014 $\pm$ 0.008 & $3.9 \pm 2.2  \times 10^{41}$ \\
PSZ2G045.50-37.21 & 0.12 & 2.81 & 321.32 & -6.98 & 0.119 $\pm$ 0.015 & 1.201 $\pm$ 0.192 & 0.003 $\pm$ 0.005 & $1.8 \pm 2.0  \times 10^{41}$ \\
PSZ2G047.73-60.15 & 0.12 & 3.36 & 342.76 & -16.40 & 0.092 $\pm$ 0.006 & 0.373 $\pm$ 0.022 & 0.008 $\pm$ 0.008 & $5.3 \pm 2.5  \times 10^{41}$ \\
PSZ2G048.75+53.18 & 0.10 & 2.53 & 234.97 & 30.70 & 0.138 $\pm$ 0.004 & 0.459 $\pm$ 0.014 & 0.004 $\pm$ 0.012 & $5.1 \pm 2.3  \times 10^{40}$ \\
PSZ2G049.32+44.37 & 0.10 & 3.67 & 245.14 & 29.89 & 0.051 $\pm$ 0.003 & 0.486 $\pm$ 0.022 & 0.043 $\pm$ 0.011 & $2.8 \pm 2.0  \times 10^{40}$ \\
PSZ2G049.69-49.46 & 0.10 & 3.63 & 333.62 & -10.35 & 0.101 $\pm$ 0.004 & 0.607 $\pm$ 0.014 & 0.004 $\pm$ 0.014 & $2.1 \pm 1.1  \times 10^{41}$ \\
PSZ2G059.72-46.21 & 0.09 & 2.23 & 334.65 & -2.96 & 0.018 $\pm$ 0.003 & 0.172 $\pm$ 0.012 & 0.025 $\pm$ 0.008 & $1.2 \pm 0.8  \times 10^{41}$ \\
PSZ2G062.44-46.43 & 0.09 & 3.47 & 335.99 & -1.62 & 0.092 $\pm$ 0.002 & 0.291 $\pm$ 0.004 & 0.046 $\pm$ 0.009 & $2.4 \pm 0.5  \times 10^{41}$ \\
PSZ2G065.32-64.84 & 0.09 & 2.48 & 351.32 & -12.13 & 0.285 $\pm$ 0.007 & 0.761 $\pm$ 0.024 & 0.004 $\pm$ 0.014 & $2.6 \pm 1.2  \times 10^{41}$ \\
PSZ2G065.81-83.77 & 0.11 & 2.83 & 7.15 & -23.63 & 0.040 $\pm$ 0.007 & 0.236 $\pm$ 0.031 & 0.018 $\pm$ 0.005 & $1.6 \pm 1.7  \times 10^{41}$ \\
PSZ2G072.40-78.45 & 0.09 & 2.76 & 3.43 & -19.49 & 0.052 $\pm$ 0.009 & 0.731 $\pm$ 0.108 & 0.005 $\pm$ 0.009 & $6.0 \pm 4.6  \times 10^{41}$ \\
PSZ2G080.41-33.24 & 0.11 & 3.77 & 336.54 & 17.38 & 0.054 $\pm$ 0.001 & 0.510 $\pm$ 0.007 & 0.039 $\pm$ 0.011 & $4.9 \pm 0.5  \times 10^{41}$ \\
PSZ2G083.14+66.57 & 0.09 & 2.07 & 213.44 & 43.65 & 0.260 $\pm$ 0.012 & 0.627 $\pm$ 0.044 & 0.004 $\pm$ 0.009 & $2.3 \pm 0.6  \times 10^{42}$ \\
PSZ2G091.40-51.01 & 0.10 & 2.68 & 353.50 & 7.06 & 0.015 $\pm$ 0.008 & 0.219 $\pm$ 0.159 & 0.016 $\pm$ 0.005 & $3.3 \pm 3.8  \times 10^{41}$ \\
PSZ2G099.57-58.64 & 0.09 & 2.24 & 0.95 & 2.05 & 0.113 $\pm$ 0.007 & 0.749 $\pm$ 0.057 & 0.007 $\pm$ 0.010 & $9.1 \pm 6.8  \times 10^{40}$ \\
PSZ2G113.29-29.69 & 0.11 & 3.71 & 2.94 & 32.43 & 0.045 $\pm$ 0.003 & 0.366 $\pm$ 0.013 & 0.008 $\pm$ 0.012 & $1.0 \pm 0.8  \times 10^{41}$ \\
PSZ2G114.79-33.71 & 0.09 & 3.82 & 5.15 & 28.67 & 0.046 $\pm$ 0.003 & 0.286 $\pm$ 0.023 & 0.016 $\pm$ 0.014 & $6.6 \pm 4.0  \times 10^{40}$ \\
PSZ2G114.90-34.35 & 0.09 & 2.47 & 5.35 & 28.05 & 0.039 $\pm$ 0.012 & -0.488 $\pm$ -0.283 & 0.025 $\pm$ 0.003 & $1.6 \pm 1.8  \times 10^{41}$ \\
PSZ2G135.19+57.88 & 0.10 & 2.21 & 180.54 & 58.06 & 0.057 $\pm$ 0.004 & 0.480 $\pm$ 0.027 & 0.005 $\pm$ 0.007 & $4.4 \pm 1.6  \times 10^{41}$ \\
PSZ2G137.74-27.08 & 0.09 & 2.83 & 28.78 & 33.94 & 0.039 $\pm$ 0.005 & 0.347 $\pm$ 0.106 & 0.030 $\pm$ 0.010 & $7.9 \pm 8.5  \times 10^{40}$ \\
PSZ2G146.00-49.45 & 0.10 & 3.88 & 27.89 & 10.71 & 0.088 $\pm$ 0.082 & 0.166 $\pm$ 0.177 & 0.031 $\pm$ 0.007 & $1.1 \pm 1.2  \times 10^{41}$ \\
PSZ2G147.11+21.84 & 0.11 & 3.38 & 94.63 & 67.41 & 0.064 $\pm$ 0.013 & 0.242 $\pm$ 0.037 & 0.012 $\pm$ 0.005 & $4.9 \pm 6.0  \times 10^{41}$ \\
PSZ2G152.70+25.46 & 0.10 & 2.95 & 106.10 & 63.32 & 0.086 $\pm$ 0.009 & 0.369 $\pm$ 0.034 & 0.008 $\pm$ 0.008 & $1.7 \pm 1.8  \times 10^{41}$ \\
PSZ2G162.33+25.03 & 0.10 & 2.85 & 108.56 & 54.69 & 0.045 $\pm$ 0.009 & 0.264 $\pm$ 0.036 & 0.004 $\pm$ 0.007 & $5.0 \pm 5.6  \times 10^{41}$ \\
PSZ2G178.94+56.00 & 0.09 & 2.23 & 154.95 & 41.01 & 0.066 $\pm$ 0.010 & -0.850 $\pm$ -0.392 & 0.013 $\pm$ 0.007 & $1.4 \pm 1.2  \times 10^{41}$ \\
PSZ2G192.18+56.12 & 0.12 & 3.62 & 154.09 & 33.66 & 0.045 $\pm$ 0.004 & 0.252 $\pm$ 0.019 & 0.003 $\pm$ 0.006 & $2.3 \pm 1.5  \times 10^{41}$ \\
PSZ2G198.47-23.69 & 0.12 & 3.63 & 75.34 & 1.16 & 0.080 $\pm$ 0.012 & 19.310 $\pm$ 5.512 & 0.006 $\pm$ 0.005 & $1.8 \pm 1.9  \times 10^{41}$ \\
PSZ2G200.63-64.71 & 0.09 & 2.48 & 39.11 & -19.35 & 0.043 $\pm$ 0.007 & 0.718 $\pm$ 0.094 & 0.032 $\pm$ 0.008 & $2.7 \pm 2.1  \times 10^{41}$ \\
PSZ2G206.22+24.89 & 0.09 & 2.27 & 122.77 & 16.63 & 0.099 $\pm$ 0.008 & 0.691 $\pm$ 0.140 & 0.014 $\pm$ 0.009 & $3.9 \pm 2.9  \times 10^{40}$ \\
PSZ2G224.38-24.95 & 0.09 & 2.54 & 84.58 & -20.65 & 0.055 $\pm$ 0.006 & 0.364 $\pm$ 0.066 & 0.004 $\pm$ 0.007 & $7.5 \pm 8.0  \times 10^{40}$ \\
PSZ2G226.16-21.95 & 0.10 & 3.87 & 88.21 & -21.07 & 0.036 $\pm$ 0.002 & 0.383 $\pm$ 0.015 & 0.014 $\pm$ 0.010 & $4.3 \pm 3.2  \times 10^{40}$ \\
PSZ2G228.62+68.44 & 0.10 & 2.91 & 170.80 & 19.61 & 0.097 $\pm$ 0.006 & 0.799 $\pm$ 0.062 & 0.001 $\pm$ 0.009 & $-3.9 \pm 4.0  \times 10^{40}$ \\
PSZ2G261.88+62.85 & 0.10 & 2.41 & 175.34 & 5.71 & 0.098 $\pm$ 0.016 & -0.664 $\pm$ -0.468 & 0.010 $\pm$ 0.007 & $2.7 \pm 2.2  \times 10^{41}$ \\
PSZ2G268.30+28.89 & 0.12 & 2.83 & 159.60 & -24.88 & 0.029 $\pm$ 0.006 & 0.293 $\pm$ 0.023 & 0.011 $\pm$ 0.004 & $3.8 \pm 3.5  \times 10^{41}$ \\
PSZ2G277.38+47.07 & 0.12 & 2.81 & 175.37 & -12.30 & 0.261 $\pm$ 0.009 & 0.615 $\pm$ 0.036 & 0.005 $\pm$ 0.007 & $4.8 \pm 1.6  \times 10^{41}$ \\
PSZ2G283.91+73.87 & 0.09 & 2.99 & 187.56 & 11.80 & 0.047 $\pm$ 0.003 & -0.561 $\pm$ -0.205 & 0.040 $\pm$ 0.015 & $2.1 \pm 0.5  \times 10^{41}$ \\
PSZ2G284.59+70.84 & 0.09 & 2.20 & 186.86 & 8.83 & 0.058 $\pm$ 0.010 & 1.707 $\pm$ 1.183 & 0.013 $\pm$ 0.015 & $5.3 \pm 5.6  \times 10^{40}$ \\
PSZ2G295.34+23.34 & 0.12 & 4.23 & 183.89 & -38.99 & 0.022 $\pm$ 0.002 & 0.182 $\pm$ 0.005 & 0.030 $\pm$ 0.008 & $1.7 \pm 0.8  \times 10^{41}$ \\
PSZ2G302.85-88.63 & 0.11 & 3.10 & 12.86 & -28.50 & 0.018 $\pm$ 0.005 & 0.156 $\pm$ 0.029 & 0.010 $\pm$ 0.004 & $5.8 \pm 4.1  \times 10^{41}$ \\
PSZ2G314.26-55.35 & 0.10 & 2.26 & 359.71 & -60.60 & 0.061 $\pm$ 0.002 & 0.128 $\pm$ 0.004 & 0.052 $\pm$ 0.009 & $6.4 \pm 1.3  \times 10^{40}$ \\
PSZ2G321.98-47.96 & 0.09 & 4.02 & 342.46 & -64.42 & 0.053 $\pm$ 0.001 & 0.358 $\pm$ 0.005 & 0.022 $\pm$ 0.014 & $7.0 \pm 1.1  \times 10^{41}$ \\
PSZ2G322.77+59.52 & 0.09 & 3.23 & 202.78 & -1.81 & 0.058 $\pm$ 0.003 & 0.396 $\pm$ 0.033 & 0.057 $\pm$ 0.015 & $1.8 \pm 0.6  \times 10^{41}$ \\
PSZ2G324.54-44.97 & 0.10 & 3.00 & 334.50 & -65.18 & 0.114 $\pm$ 0.004 & 0.494 $\pm$ 0.022 & 0.030 $\pm$ 0.012 & $2.4 \pm 0.8  \times 10^{41}$ \\
PSZ2G336.60-55.43 & 0.10 & 4.28 & 341.57 & -52.72 & 0.024 $\pm$ 0.002 & 0.311 $\pm$ 0.012 & 0.012 $\pm$ 0.010 & $9.0 \pm 9.6  \times 10^{40}$ \\
PSZ2G340.46-52.20 & 0.10 & 2.54 & 335.08 & -52.45 & 0.073 $\pm$ 0.006 & 0.781 $\pm$ 0.098 & 0.023 $\pm$ 0.010 & $4.1 \pm 1.3  \times 10^{41}$ \\
PSZ2G342.51-50.97 & 0.11 & 2.37 & 332.38 & -51.82 & 0.074 $\pm$ 0.008 & 0.321 $\pm$ 0.026 & 0.015 $\pm$ 0.006 & $4.9 \pm 3.1  \times 10^{41}$ \\
PSZ2G346.30-30.35 & 0.11 & 2.95 & 298.31 & -52.07 & 0.027 $\pm$ 0.009 & 0.322 $\pm$ 0.068 & 0.010 $\pm$ 0.007 & $1.4 \pm 1.7  \times 10^{41}$ \\
PSZ2G346.36-77.71 & 0.12 & 2.46 & 2.49 & -35.66 & 0.062 $\pm$ 0.011 & 0.405 $\pm$ 0.063 & 0.007 $\pm$ 0.005 & $7.9 \pm 8.7  \times 10^{40}$ \\
PSZ2G352.28-77.66 & 0.11 & 4.00 & 1.49 & -34.70 & 0.036 $\pm$ 0.007 & 0.300 $\pm$ 0.024 & 0.007 $\pm$ 0.006 & $2.7 \pm 3.1  \times 10^{41}$ \\
PSZ2G356.53+28.66 & 0.10 & 2.95 & 239.61 & -14.16 & 0.177 $\pm$ 0.006 & 0.570 $\pm$ 0.026 & 0.004 $\pm$ 0.013 & $1.6 \pm 0.5  \times 10^{42}$ \\
PSZ2G358.21-87.49 & 0.11 & 3.65 & 10.52 & -28.53 & 0.045 $\pm$ 0.005 & 0.374 $\pm$ 0.017 & 0.010 $\pm$ 0.007 & $6.4 \pm 5.8  \times 10^{41}$ \\
PSZ2G358.94-70.57 & 0.10 & 2.52 & 352.78 & -36.55 & 0.191 $\pm$ 0.008 & 0.630 $\pm$ 0.036 & 0.003 $\pm$ 0.008 & $2.3 \pm 1.1  \times 10^{41}$ \\

& & & & & & & & \\

PSZ2G000.13+78.04 & 0.17 & 5.12 & 203.56 & 20.26 & 0.025 $\pm$ 0.004 & 0.270 $\pm$ 0.017 & 0.016 $\pm$ 0.007 & $5.8 \pm 3.7  \times 10^{41}$ \\
PSZ2G000.40-41.86 & 0.17 & 5.30 & 316.08 & -41.35 & 0.050 $\pm$ 0.003 & 0.447 $\pm$ 0.014 & 0.018 $\pm$ 0.009 & $1.5 \pm 0.9  \times 10^{41}$ \\
PSZ2G002.82+39.23 & 0.15 & 5.74 & 235.01 & -3.29 & 0.043 $\pm$ 0.006 & 0.457 $\pm$ 0.035 & 0.012 $\pm$ 0.011 & $3.8 \pm 4.0  \times 10^{41}$ \\
PSZ2G003.91-42.03 & 0.15 & 4.72 & 316.47 & -38.75 & 0.030 $\pm$ 0.010 & 0.313 $\pm$ 0.042 & 0.060 $\pm$ 0.009 & $4.3 \pm 5.0  \times 10^{41}$ \\
PSZ2G003.93-59.41 & 0.15 & 7.19 & 338.61 & -37.74 & 0.034 $\pm$ 0.001 & 0.449 $\pm$ 0.007 & 0.022 $\pm$ 0.012 & $2.9 \pm 1.5  \times 10^{40}$ \\
PSZ2G006.38+62.03 & 0.24 & 4.81 & 218.31 & 12.49 & 0.098 $\pm$ 0.027 & -0.334 $\pm$ -0.098 & 0.005 $\pm$ 0.004 & $7.5 \pm 6.5  \times 10^{41}$ \\
PSZ2G014.09+38.38 & 0.22 & 4.73 & 240.86 & 3.32 & 0.054 $\pm$ 0.006 & 0.384 $\pm$ 0.023 & 0.010 $\pm$ 0.004 & $9.3 \pm 9.6  \times 10^{40}$ \\
PSZ2G018.32-28.50 & 0.16 & 5.09 & 303.70 & -24.51 & 0.312 $\pm$ 0.007 & 0.725 $\pm$ 0.017 & 0.004 $\pm$ 0.012 & $1.9 \pm 0.4  \times 10^{42}$ \\
PSZ2G021.10+33.24 & 0.15 & 7.79 & 248.20 & 5.59 & 0.297 $\pm$ 0.001 & 0.661 $\pm$ 0.002 & 0.001 $\pm$ 0.021 & $4.5 \pm 0.1  \times 10^{42}$ \\
PSZ2G024.44+22.76 & 0.16 & 4.51 & 258.83 & 3.17 & 0.157 $\pm$ 0.007 & 0.608 $\pm$ 0.021 & 0.004 $\pm$ 0.009 & $1.5 \pm 0.6  \times 10^{42}$ \\
PSZ2G026.73-38.94 & 0.20 & 5.41 & 316.84 & -21.18 & 0.029 $\pm$ 0.011 & -1.778 $\pm$ -0.495 & 0.018 $\pm$ 0.003 & $4.8 \pm 6.8  \times 10^{41}$ \\
PSZ2G028.89+60.13 & 0.15 & 4.54 & 225.09 & 21.36 & 0.178 $\pm$ 0.003 & 0.673 $\pm$ 0.009 & 0.003 $\pm$ 0.012 & $1.4 \pm 0.2  \times 10^{42}$ \\
PSZ2G033.97-76.61 & 0.23 & 7.56 & 357.93 & -26.10 & 0.155 $\pm$ 0.007 & 0.714 $\pm$ 0.035 & 0.007 $\pm$ 0.009 & $6.9 \pm 2.1  \times 10^{42}$ \\
PSZ2G039.85-39.96 & 0.18 & 5.92 & 321.75 & -12.16 & 0.030 $\pm$ 0.001 & 0.109 $\pm$ 0.001 & 0.043 $\pm$ 0.008 & $3.2 \pm 1.2  \times 10^{40}$ \\
PSZ2G041.45+29.10 & 0.18 & 5.59 & 259.47 & 19.69 & 0.017 $\pm$ 0.005 & 0.153 $\pm$ 0.016 & 0.014 $\pm$ 0.005 & $4.6 \pm 5.0  \times 10^{41}$ \\
PSZ2G045.13+67.78 & 0.22 & 4.83 & 218.00 & 29.56 & 0.057 $\pm$ 0.013 & 0.198 $\pm$ 0.035 & 0.009 $\pm$ 0.004 & $2.8 \pm 3.0  \times 10^{41}$ \\
PSZ2G047.71-59.47 & 0.25 & 4.61 & 342.12 & -16.11 & 0.074 $\pm$ 0.004 & 0.395 $\pm$ 0.015 & 0.004 $\pm$ 0.006 & $-1.0 \pm 0.6  \times 10^{41}$ \\
PSZ2G049.18+65.05 & 0.23 & 4.73 & 221.12 & 31.23 & 0.194 $\pm$ 0.018 & 0.582 $\pm$ 0.088 & 0.009 $\pm$ 0.005 & $1.8 \pm 0.8  \times 10^{42}$ \\
PSZ2G049.22+30.87 & 0.16 & 5.90 & 260.02 & 26.63 & 0.213 $\pm$ 0.002 & 0.627 $\pm$ 0.005 & 0.002 $\pm$ 0.013 & $6.6 \pm 0.6  \times 10^{41}$ \\
PSZ2G054.99+53.41 & 0.23 & 5.73 & 234.90 & 34.43 & 0.037 $\pm$ 0.002 & 0.288 $\pm$ 0.006 & 0.023 $\pm$ 0.008 & $4.3 \pm 1.6  \times 10^{41}$ \\
PSZ2G055.59+31.85 & 0.22 & 7.78 & 260.62 & 32.14 & 0.113 $\pm$ 0.002 & 0.502 $\pm$ 0.005 & 0.006 $\pm$ 0.011 & $4.2 \pm 0.8  \times 10^{41}$ \\
PSZ2G055.95-34.89 & 0.23 & 6.73 & 323.83 & 1.41 & 0.021 $\pm$ 0.003 & 0.173 $\pm$ 0.006 & 0.024 $\pm$ 0.007 & $5.0 \pm 2.7  \times 10^{41}$ \\
PSZ2G057.08-74.45 & 0.25 & 4.71 & 357.78 & -19.98 & 0.074 $\pm$ 0.011 & 0.154 $\pm$ 0.013 & 0.022 $\pm$ 0.004 & $1.5 \pm 1.5  \times 10^{41}$ \\
PSZ2G057.73+51.58 & 0.24 & 5.59 & 237.14 & 36.10 & 0.026 $\pm$ 0.010 & 29.943 $\pm$ 6.604 & 0.007 $\pm$ 0.005 & $1.1 \pm 1.0  \times 10^{42}$ \\
PSZ2G059.81-39.09 & 0.22 & 4.99 & 329.04 & 1.42 & 0.041 $\pm$ 0.006 & 0.286 $\pm$ 0.017 & 0.012 $\pm$ 0.003 & $5.8 \pm 4.3  \times 10^{41}$ \\
PSZ2G067.17+67.46 & 0.17 & 7.24 & 216.50 & 37.82 & 0.065 $\pm$ 0.001 & 0.478 $\pm$ 0.003 & 0.011 $\pm$ 0.011 & $2.0 \pm 0.2  \times 10^{41}$ \\
PSZ2G067.52+34.75 & 0.18 & 4.54 & 259.31 & 42.46 & 0.266 $\pm$ 0.012 & 0.961 $\pm$ 0.057 & 0.007 $\pm$ 0.006 & $2.3 \pm 0.6  \times 10^{42}$ \\
PSZ2G070.08-31.79 & 0.19 & 4.96 & 328.92 & 12.52 & 0.109 $\pm$ 0.005 & 0.523 $\pm$ 0.022 & 0.001 $\pm$ 0.008 & $1.7 \pm 0.6  \times 10^{42}$ \\
PSZ2G073.97-27.82 & 0.23 & 9.80 & 328.42 & 17.70 & 0.087 $\pm$ 0.007 & 0.499 $\pm$ 0.017 & 0.006 $\pm$ 0.009 & $6.2 \pm 2.7  \times 10^{42}$ \\
PSZ2G074.75-24.59 & 0.25 & 5.03 & 326.54 & 20.48 & 0.094 $\pm$ 0.010 & 0.541 $\pm$ 0.048 & 0.003 $\pm$ 0.005 & $1.9 \pm 0.9  \times 10^{42}$ \\
PSZ2G083.86+85.09 & 0.18 & 4.66 & 196.46 & 30.89 & 0.054 $\pm$ 0.005 & 0.346 $\pm$ 0.020 & 0.011 $\pm$ 0.006 & 0 \\
PSZ2G092.16-66.01 & 0.23 & 5.96 & 0.79 & -6.07 & 0.039 $\pm$ 0.007 & 0.402 $\pm$ 0.024 & 0.006 $\pm$ 0.006 & $1.5 \pm 1.6  \times 10^{42}$ \\
PSZ2G092.71+73.46 & 0.23 & 8.13 & 203.80 & 41.00 & 0.036 $\pm$ 0.001 & 0.295 $\pm$ 0.003 & 0.007 $\pm$ 0.007 & $7.0 \pm 2.1  \times 10^{40}$ \\
PSZ2G097.72+38.12 & 0.17 & 6.59 & 248.98 & 66.20 & 0.034 $\pm$ 0.001 & 0.321 $\pm$ 0.003 & 0.017 $\pm$ 0.007 & $1.1 \pm 0.3  \times 10^{41}$ \\
PSZ2G101.52-29.98 & 0.23 & 4.88 & 351.60 & 29.33 & 0.090 $\pm$ 0.010 & 0.830 $\pm$ 0.062 & 0.004 $\pm$ 0.004 & $4.4 \pm 1.7  \times 10^{42}$ \\
PSZ2G104.71-54.54 & 0.24 & 4.90 & 2.34 & 6.83 & 0.044 $\pm$ 0.008 & 0.186 $\pm$ 0.019 & 0.012 $\pm$ 0.005 & $3.7 \pm 2.9  \times 10^{41}$ \\
PSZ2G113.81+44.35 & 0.23 & 4.63 & 213.53 & 71.28 & 0.029 $\pm$ 0.003 & 0.192 $\pm$ 0.007 & 0.045 $\pm$ 0.005 & $1.0 \pm 0.7  \times 10^{41}$ \\
PSZ2G114.99+70.36 & 0.23 & 5.70 & 196.73 & 46.53 & 0.034 $\pm$ 0.002 & 0.233 $\pm$ 0.006 & 0.019 $\pm$ 0.007 & $1.1 \pm 0.6  \times 10^{41}$ \\
PSZ2G118.58+28.57 & 0.18 & 5.98 & 261.23 & 85.89 & 0.056 $\pm$ 0.003 & 0.366 $\pm$ 0.010 & 0.010 $\pm$ 0.008 & $1.8 \pm 0.8  \times 10^{41}$ \\
PSZ2G124.20-36.48 & 0.20 & 7.65 & 14.00 & 26.38 & 0.151 $\pm$ 0.001 & 0.228 $\pm$ 0.001 & 0.073 $\pm$ 0.009 & $5.5 \pm 0.3  \times 10^{41}$ \\
PSZ2G125.30-27.99 & 0.22 & 5.02 & 15.41 & 34.83 & 0.030 $\pm$ 0.009 & 0.823 $\pm$ 0.111 & 0.012 $\pm$ 0.004 & $4.2 \pm 4.8  \times 10^{41}$ \\
PSZ2G127.44-34.74 & 0.25 & 4.90 & 17.06 & 27.98 & 0.025 $\pm$ 0.006 & 0.152 $\pm$ 0.015 & 0.032 $\pm$ 0.003 & $7.8 \pm 6.2  \times 10^{41}$ \\
PSZ2G140.63+29.45 & 0.21 & 5.02 & 115.52 & 74.25 & 0.192 $\pm$ 0.001 & 0.482 $\pm$ 0.002 & 0.003 $\pm$ 0.006 & $9.3 \pm 0.3  \times 10^{41}$ \\
PSZ2G143.62+42.61 & 0.21 & 4.95 & 150.84 & 67.14 & 0.012 $\pm$ 0.006 & 0.170 $\pm$ 0.017 & 0.017 $\pm$ 0.005 & $4.2 \pm 4.8  \times 10^{41}$ \\
PSZ2G145.19+32.14 & 0.22 & 4.87 & 122.77 & 70.05 & 0.074 $\pm$ 0.007 & 0.480 $\pm$ 0.022 & 0.014 $\pm$ 0.005 & $5.0 \pm 3.0  \times 10^{41}$ \\
PSZ2G149.39-36.84 & 0.17 & 5.42 & 35.38 & 21.35 & 0.033 $\pm$ 0.004 & 0.240 $\pm$ 0.015 & 0.013 $\pm$ 0.006 & $3.1 \pm 1.9  \times 10^{41}$ \\
PSZ2G149.75+34.68 & 0.18 & 8.86 & 127.72 & 65.86 & 0.055 $\pm$ 0.001 & 0.272 $\pm$ 0.002 & 0.036 $\pm$ 0.011 & $-1.0 \pm 0.2  \times 10^{41}$ \\
PSZ2G150.90-39.05 & 0.22 & 5.00 & 35.50 & 18.85 & 0.041 $\pm$ 0.009 & 0.998 $\pm$ 0.125 & 0.008 $\pm$ 0.005 & $1.1 \pm 0.9  \times 10^{42}$ \\
PSZ2G153.00-58.26 & 0.23 & 5.04 & 28.15 & 0.99 & 0.056 $\pm$ 0.003 & 0.483 $\pm$ 0.008 & 0.011 $\pm$ 0.006 & $3.3 \pm 1.2  \times 10^{41}$ \\
PSZ2G159.91-73.50 & 0.21 & 8.46 & 22.98 & -13.61 & 0.032 $\pm$ 0.002 & 0.294 $\pm$ 0.005 & 0.014 $\pm$ 0.011 & $4.0 \pm 1.6  \times 10^{41}$ \\
PSZ2G161.94-48.17 & 0.22 & 6.54 & 37.88 & 6.97 & 0.033 $\pm$ 0.006 & 0.420 $\pm$ 0.038 & 0.014 $\pm$ 0.005 & $3.7 \pm 3.9  \times 10^{41}$ \\
PSZ2G162.49-71.98 & 0.21 & 4.64 & 24.51 & -12.80 & 0.072 $\pm$ 0.003 & 0.464 $\pm$ 0.010 & 0.018 $\pm$ 0.007 & $3.7 \pm 1.0  \times 10^{41}$ \\
PSZ2G163.69+53.52 & 0.16 & 4.73 & 155.60 & 50.12 & 0.041 $\pm$ 0.002 & 0.401 $\pm$ 0.009 & 0.014 $\pm$ 0.007 & $2.7 \pm 1.2  \times 10^{41}$ \\
PSZ2G166.09+43.38 & 0.22 & 6.85 & 139.48 & 51.72 & 0.045 $\pm$ 0.001 & 0.367 $\pm$ 0.004 & 0.006 $\pm$ 0.006 & $4.6 \pm 1.2  \times 10^{41}$ \\
PSZ2G166.62+42.13 & 0.23 & 5.34 & 137.40 & 51.55 & 0.011 $\pm$ 0.004 & 0.024 $\pm$ 0.002 & 0.081 $\pm$ 0.006 & $1.7 \pm 0.5  \times 10^{42}$ \\
PSZ2G175.69-85.98 & 0.23 & 5.67 & 16.38 & -24.65 & 0.005 $\pm$ 0.001 & 0.180 $\pm$ 0.005 & 0.017 $\pm$ 0.007 & $7.1 \pm 2.2  \times 10^{41}$ \\
PSZ2G176.25-52.57 & 0.24 & 6.07 & 42.06 & -2.24 & 0.033 $\pm$ 0.004 & 0.323 $\pm$ 0.019 & 0.010 $\pm$ 0.005 & $8.7 \pm 6.5  \times 10^{41}$ \\
PSZ2G180.60+76.65 & 0.21 & 6.30 & 179.31 & 33.61 & 0.119 $\pm$ 0.003 & 0.465 $\pm$ 0.006 & 0.009 $\pm$ 0.009 & $6.4 \pm 1.3  \times 10^{41}$ \\
PSZ2G182.59+55.83 & 0.21 & 5.83 & 154.26 & 39.03 & 0.104 $\pm$ 0.005 & 0.445 $\pm$ 0.014 & 0.002 $\pm$ 0.010 & $1.5 \pm 0.5  \times 10^{42}$ \\
PSZ2G187.53+21.92 & 0.17 & 5.17 & 113.06 & 31.63 & 0.089 $\pm$ 0.001 & 0.515 $\pm$ 0.005 & 0.005 $\pm$ 0.010 & $8.4 \pm 0.9  \times 10^{40}$ \\
PSZ2G189.53-25.10 & 0.24 & 6.04 & 69.77 & 7.27 & 0.100 $\pm$ 0.003 & 0.499 $\pm$ 0.007 & 0.006 $\pm$ 0.007 & $7.9 \pm 2.0  \times 10^{41}$ \\
PSZ2G195.75-24.32 & 0.20 & 7.80 & 73.53 & 2.96 & 0.026 $\pm$ 0.000 & 0.146 $\pm$ 0.001 & 0.049 $\pm$ 0.008 & $3.1 \pm 0.4  \times 10^{40}$ \\
PSZ2G200.95-28.16 & 0.22 & 5.30 & 72.58 & -3.01 & 0.018 $\pm$ 0.019 & 0.078 $\pm$ 0.005 & 0.028 $\pm$ 0.004 & $1.2 \pm 1.3  \times 10^{41}$ \\
PSZ2G208.60-26.00 & 0.22 & 7.74 & 77.69 & -8.02 & 0.049 $\pm$ 0.002 & 0.319 $\pm$ 0.006 & 0.022 $\pm$ 0.009 & $2.5 \pm 1.1  \times 10^{41}$ \\
PSZ2G208.80-30.67 & 0.25 & 7.26 & 73.52 & -10.23 & 0.030 $\pm$ 0.002 & 0.143 $\pm$ 0.003 & 0.056 $\pm$ 0.006 & $1.0 \pm 0.4  \times 10^{41}$ \\
PSZ2G212.82-84.04 & 0.23 & 5.68 & 19.55 & -26.96 & 0.041 $\pm$ 0.002 & 0.335 $\pm$ 0.007 & 0.021 $\pm$ 0.006 & $4.3 \pm 1.8  \times 10^{41}$ \\
PSZ2G214.61-22.70 & 0.15 & 5.39 & 83.10 & -11.54 & 0.045 $\pm$ 0.005 & 0.342 $\pm$ 0.013 & 0.011 $\pm$ 0.010 & $3.0 \pm 3.1  \times 10^{41}$ \\
PSZ2G215.19-49.65 & 0.24 & 5.99 & 57.00 & -21.74 & 0.022 $\pm$ 0.004 & 0.195 $\pm$ 0.012 & 0.025 $\pm$ 0.005 & $2.5 \pm 1.9  \times 10^{41}$ \\
PSZ2G215.61+22.61 & 0.25 & 5.19 & 124.29 & 7.87 & 0.052 $\pm$ 0.007 & 0.660 $\pm$ 0.045 & 0.007 $\pm$ 0.003 & $1.0 \pm 1.0  \times 10^{41}$ \\
PSZ2G218.81+35.51 & 0.18 & 5.42 & 137.29 & 10.99 & 0.083 $\pm$ 0.002 & 0.369 $\pm$ 0.005 & 0.032 $\pm$ 0.011 & $1.4 \pm 0.4  \times 10^{41}$ \\
PSZ2G219.88+22.83 & 0.23 & 5.07 & 126.28 & 4.46 & 0.021 $\pm$ 0.008 & 0.115 $\pm$ 0.017 & 0.023 $\pm$ 0.005 & $4.1 \pm 4.5  \times 10^{41}$ \\
PSZ2G220.11+22.91 & 0.22 & 4.87 & 126.45 & 4.30 & 0.103 $\pm$ 0.005 & 0.526 $\pm$ 0.017 & 0.027 $\pm$ 0.008 & $1.2 \pm 0.3  \times 10^{42}$ \\
PSZ2G222.99-65.70 & 0.23 & 6.18 & 40.35 & -28.66 & 0.038 $\pm$ 0.005 & 0.324 $\pm$ 0.023 & 0.011 $\pm$ 0.007 & $1.2 \pm 0.6  \times 10^{42}$ \\
PSZ2G224.00+69.33 & 0.19 & 5.27 & 171.00 & 21.49 & 0.046 $\pm$ 0.003 & 0.296 $\pm$ 0.010 & 0.007 $\pm$ 0.007 & 0 \\
PSZ2G225.48+29.41 & 0.20 & 4.60 & 134.47 & 3.17 & 0.043 $\pm$ 0.006 & 0.433 $\pm$ 0.027 & 0.006 $\pm$ 0.006 & $1.8 \pm 1.9  \times 10^{41}$ \\
PSZ2G226.75+48.95 & 0.22 & 4.85 & 152.22 & 11.79 & 0.042 $\pm$ 0.002 & 0.275 $\pm$ 0.005 & 0.019 $\pm$ 0.005 & $1.6 \pm 0.5  \times 10^{41}$ \\
PSZ2G244.71+32.50 & 0.15 & 5.23 & 146.38 & -8.64 & 0.031 $\pm$ 0.002 & 0.385 $\pm$ 0.012 & 0.003 $\pm$ 0.009 & $2.3 \pm 1.7  \times 10^{40}$ \\
PSZ2G249.38+33.26 & 0.17 & 5.41 & 149.60 & -11.07 & 0.150 $\pm$ 0.001 & 0.534 $\pm$ 0.003 & 0.002 $\pm$ 0.010 & $7.8 \pm 0.6  \times 10^{41}$ \\
PSZ2G250.89-36.24 & 0.20 & 5.74 & 77.60 & -45.33 & 0.061 $\pm$ 0.003 & 0.450 $\pm$ 0.009 & 0.005 $\pm$ 0.007 & $2.8 \pm 1.7  \times 10^{40}$ \\
PSZ2G253.13+57.49 & 0.23 & 4.84 & 168.56 & 4.35 & 0.051 $\pm$ 0.008 & 0.671 $\pm$ 0.065 & 0.006 $\pm$ 0.004 & $1.7 \pm 1.8  \times 10^{41}$ \\
PSZ2G253.48-33.72 & 0.19 & 4.67 & 81.46 & -47.26 & 0.075 $\pm$ 0.003 & 0.469 $\pm$ 0.008 & 0.004 $\pm$ 0.006 & $5.5 \pm 1.9  \times 10^{41}$ \\
PSZ2G256.53-65.70 & 0.22 & 6.08 & 36.46 & -41.92 & 0.107 $\pm$ 0.003 & 0.425 $\pm$ 0.006 & 0.005 $\pm$ 0.008 & $2.4 \pm 0.4  \times 10^{42}$ \\
PSZ2G257.32-22.19 & 0.20 & 5.31 & 99.33 & -48.48 & 0.076 $\pm$ 0.004 & 0.284 $\pm$ 0.007 & 0.048 $\pm$ 0.006 & $3.0 \pm 1.2  \times 10^{41}$ \\
PSZ2G263.14-23.41 & 0.23 & 6.83 & 99.72 & -53.98 & 0.097 $\pm$ 0.001 & 0.443 $\pm$ 0.002 & 0.005 $\pm$ 0.009 & $7.6 \pm 0.7  \times 10^{41}$ \\
PSZ2G263.68-22.55 & 0.16 & 7.96 & 101.37 & -54.23 & 0.072 $\pm$ 0.003 & 0.511 $\pm$ 0.009 & 0.009 $\pm$ 0.010 & $6.3 \pm 2.3  \times 10^{41}$ \\
PSZ2G280.17+47.83 & 0.16 & 6.06 & 177.45 & -12.30 & 0.030 $\pm$ 0.002 & 0.284 $\pm$ 0.008 & 0.022 $\pm$ 0.009 & $2.1 \pm 1.0  \times 10^{41}$ \\
PSZ2G285.63+72.75 & 0.17 & 5.61 & 187.72 & 10.58 & 0.037 $\pm$ 0.002 & 0.435 $\pm$ 0.011 & 0.015 $\pm$ 0.008 & $1.6 \pm 1.0  \times 10^{41}$ \\
PSZ2G286.62-31.24 & 0.21 & 5.68 & 82.84 & -75.19 & 0.033 $\pm$ 0.002 & 0.302 $\pm$ 0.007 & 0.020 $\pm$ 0.007 & $2.6 \pm 1.4  \times 10^{41}$ \\
PSZ2G287.96-32.99 & 0.25 & 6.18 & 74.85 & -75.80 & 0.028 $\pm$ 0.015 & 0.260 $\pm$ 0.021 & 0.025 $\pm$ 0.005 & $4.8 \pm 6.8  \times 10^{42}$ \\
PSZ2G288.27+39.97 & 0.20 & 7.16 & 180.82 & -21.52 & 0.034 $\pm$ 0.006 & 0.363 $\pm$ 0.025 & 0.014 $\pm$ 0.007 & $5.2 \pm 5.5  \times 10^{41}$ \\
PSZ2G289.13+72.19 & 0.23 & 5.88 & 188.62 & 9.78 & 0.029 $\pm$ 0.003 & 0.145 $\pm$ 0.005 & 0.026 $\pm$ 0.007 & $3.8 \pm 2.0  \times 10^{41}$ \\
PSZ2G304.65-31.66 & 0.19 & 4.71 & 354.95 & -85.22 & 0.027 $\pm$ 0.003 & 0.139 $\pm$ 0.005 & 0.019 $\pm$ 0.005 & $2.9 \pm 1.6  \times 10^{41}$ \\
PSZ2G306.02+34.47 & 0.22 & 5.00 & 195.75 & -28.33 & 0.045 $\pm$ 0.016 & 0.701 $\pm$ 0.076 & 0.010 $\pm$ 0.003 & $1.9 \pm 2.2  \times 10^{42}$ \\
PSZ2G309.49+37.31 & 0.24 & 6.72 & 198.63 & -25.27 & 0.031 $\pm$ 0.004 & 0.222 $\pm$ 0.011 & 0.016 $\pm$ 0.005 & $8.8 \pm 6.5  \times 10^{41}$ \\
PSZ2G313.33+61.13 & 0.18 & 8.77 & 197.86 & -1.34 & 0.122 $\pm$ 0.001 & 0.587 $\pm$ 0.002 & 0.001 $\pm$ 0.016 & $3.5 \pm 0.2  \times 10^{41}$ \\
PSZ2G315.56-37.04 & 0.23 & 5.69 & 328.91 & -75.43 & 0.046 $\pm$ 0.006 & 0.391 $\pm$ 0.022 & 0.012 $\pm$ 0.004 & $1.4 \pm 0.9  \times 10^{42}$ \\
PSZ2G318.14-29.57 & 0.22 & 5.89 & 296.78 & -76.40 & 0.135 $\pm$ 0.004 & 0.486 $\pm$ 0.007 & 0.006 $\pm$ 0.007 & $3.4 \pm 0.5  \times 10^{42}$ \\
PSZ2G322.63-49.15 & 0.21 & 4.65 & 343.53 & -63.25 & 0.128 $\pm$ 0.009 & 0.515 $\pm$ 0.023 & 0.007 $\pm$ 0.006 & $2.6 \pm 1.0  \times 10^{42}$ \\
PSZ2G340.94+35.07 & 0.24 & 7.80 & 224.89 & -18.17 & 0.242 $\pm$ 0.011 & 0.685 $\pm$ 0.028 & 0.004 $\pm$ 0.009 & $1.6 \pm 0.4  \times 10^{43}$ \\
PSZ2G341.44-40.19 & 0.25 & 4.77 & 315.26 & -55.72 & 0.056 $\pm$ 0.020 & 1.211 $\pm$ 0.177 & 0.011 $\pm$ 0.004 & $2.7 \pm 2.7  \times 10^{42}$ \\
PSZ2G342.33-34.93 & 0.23 & 6.68 & 305.85 & -55.58 & 0.033 $\pm$ 0.001 & 0.231 $\pm$ 0.002 & 0.022 $\pm$ 0.005 & $1.4 \pm 0.4  \times 10^{41}$ \\
PSZ2G346.61+35.06 & 0.22 & 8.86 & 228.75 & -15.36 & 0.017 $\pm$ 0.001 & 0.136 $\pm$ 0.003 & 0.053 $\pm$ 0.008 & $2.0 \pm 0.9  \times 10^{41}$ \\
PSZ2G347.17-27.36 & 0.24 & 7.19 & 293.72 & -50.88 & 0.024 $\pm$ 0.002 & 0.214 $\pm$ 0.006 & 0.010 $\pm$ 0.007 & $4.2 \pm 3.0  \times 10^{40}$ \\
PSZ2G348.43-25.50 & 0.25 & 5.81 & 291.24 & -49.41 & 0.033 $\pm$ 0.015 & 0.473 $\pm$ 0.053 & 0.015 $\pm$ 0.002 & $2.3 \pm 2.8  \times 10^{42}$ \\
PSZ2G350.12+45.29 & 0.22 & 5.31 & 224.11 & -5.82 & 0.034 $\pm$ 0.006 & 0.268 $\pm$ 0.018 & 0.016 $\pm$ 0.006 & $4.0 \pm 4.3  \times 10^{41}$ \\
PSZ2G355.07+46.20 & 0.22 & 6.67 & 226.02 & -2.80 & 0.322 $\pm$ 0.001 & 0.750 $\pm$ 0.002 & 0.001 $\pm$ 0.015 & $4.7 \pm 0.1  \times 10^{42}$ \\
PSZ2G357.43+69.50 & 0.21 & 5.66 & 209.97 & 14.25 & 0.061 $\pm$ 0.007 & 0.456 $\pm$ 0.022 & 0.011 $\pm$ 0.006 & $7.1 \pm 5.3  \times 10^{41}$ \\
PSZ2G358.98-67.26 & 0.18 & 5.22 & 348.93 & -37.78 & 0.068 $\pm$ 0.008 & 0.397 $\pm$ 0.023 & 0.031 $\pm$ 0.007 & $6.4 \pm 4.8  \times 10^{41}$ \\

\end{longtable}

\end{appendix}

\end{document}